\documentstyle[12pt,epsf]{article}
\newcommand{\be}{\begin{equation}}
\newcommand{\ee}{\end{equation}}

\newcommand{\beqa}{\begin{eqnarray}}
\newcommand{\eeqa}{\end{eqnarray}}
\newcommand{\nn}{\nonumber}

\newcommand{\eqref}[1]{(\ref{#1})}

\def\boxit#1{\vbox{\hrule\hbox{\vrule\kern8pt
\vbox{\hbox{\kern8pt}\hbox{\vbox{#1}}\hbox{\kern8pt}}
\kern8pt\vrule}\hrule}}
\def\mathboxit#1{\vbox{\hrule\hbox{\vrule\kern8pt\vbox{\kern8pt
\hbox{$\displaystyle #1$}\kern8pt}\kern8pt\vrule}\hrule}}

\def\IB{\relax\hbox{$\inbar\kern-.3em{\rm B}$}}
\def\IC{\relax\hbox{$\inbar\kern-.3em{\rm C}$}}
\def\ID{\relax\hbox{$\inbar\kern-.3em{\rm D}$}}
\def\IE{\relax\hbox{$\inbar\kern-.3em{\rm E}$}}
\def\IF{\relax\hbox{$\inbar\kern-.3em{\rm F}$}}
\def\IG{\relax\hbox{$\inbar\kern-.3em{\rm G}$}}
\def\IGa{\relax\hbox{${\rm I}\kern-.18em\Gamma$}}
\def\IH{\relax{\rm I\kern-.18em H}}
\def\IK{\relax{\rm I\kern-.18em K}}
\def\IL{\relax{\rm I\kern-.18em L}}
\def\IP{\relax{\rm I\kern-.18em P}}
\def\IR{\relax{\rm I\kern-.18em R}}
\def\IZ{\relax\ifmmode\mathchoice
{\hbox{\cmss Z\kern-.4em Z}}{\hbox{\cmss Z\kern-.4em Z}}
{\lower.9pt\hbox{\cmsss Z\kern-.4em Z}} {\lower1.2pt\hbox{\cmsss
Z\kern-.4em Z}}\else{\cmss Z\kern-.4em Z}\fi}

\def\II{\relax{\rm I\kern-.18em I}}

\begin{document}

\hfill  NRCPS-HE-05-77

\vspace{1cm}
\begin{center}
{\LARGE ~\\ Non-Abelian Tensor Gauge Fields}\\

{\large ~\\Generalization of Yang-Mills Theory

~\\

}

\vspace{1cm}

{\sl George Savvidy\\
Demokritos National Research Center\\
Institute of Nuclear Physics\\
Ag. Paraskevi, GR-15310 Athens,Greece  \\
\centerline{\footnotesize\it E-mail: savvidy@inp.demokritos.gr}
}
\end{center}
\vspace{60pt}

\centerline{{\bf Abstract}}

\vspace{12pt}

\noindent
We suggest an extension of the gauge principle which includes tensor gauge fields.
The extended non-Abelian gauge
transformations of the tensor gauge fields form a new large group.
On this group one can define field strength tensors,
which are transforming homogeneously with respect to the extended
gauge transformations.  The invariant Lagrangian is quadratic in
the field strength tensors and describes interaction of tensor gauge
fields of arbitrary large integer spin $1,2,...$. It does not contain
higher derivatives of the tensor gauge fields, and all interactions take place
through three- and four-particle exchanges with dimensionless coupling constant.
In this extension of the Yang-Mills theory the
vector gauge boson becomes a member of a bigger family of tensor gauge bosons.

We shall present a second invariant Lagrangian which can be
constructed in terms of the above field strength tensors. The
total Lagrangian is a sum of the two Lagrangians and exhibits
enhanced local gauge invariance with double number of gauge
parameters. This allows to eliminate all negative norm states of
the nonsymmetric second-rank tensor gauge field, which describes
therefore three physical polarizations: two symmetric polarizations of
helicity-two massless charged tensor gauge bosons and
antisymmetric polarization of helicity-zero charged B field.


\newpage

\pagestyle{plain}

\section{{\it Introduction}}

It is well understood, that the concept of local gauge
invariance allows to define the non-Abelian gauge fields,
to derive their  dynamical field equations
and to develop a universal point of view on matter interactions
as resulting from the exchange of gauge quanta of different forms \cite{yang,yukawa}.
The fundamental forces - electromagnetic, weak and strong interactions -
can successfully be described by the non-Abelian Yang-Mills fields.
The vector-like gauge particles - the photon, $W^{\pm},Z$ and
gluons - mediate interaction
between smallest constituents of matter - leptons and quarks
\cite{Schwinger:1957em,Glashow:1961tr,salam,weinberg,Feynman:1963ax,Faddeev:1967fc,
DeWitt:1967ub,Mandelstam:1968hz,Slavnov:1970tk,'tHooft:1971fh,'tHooft:1971rn,'tHooft:1972fi,
Lee:1971kj,Lee:1972fj,Slavnov:1972fg,Taylor:1971ff,
Gross:1973id,Politzer:1973fx,Fritzsch:1973pi,Savvidy:1977as}.

The non-Abelian local gauge invariance, which was formulated
by Yang and Mills in \cite{yang},
requires that all interactions must be invariant under
independent rotations of internal
charges at all space-time
points \footnote{The early formulation of Abelian gauge invariance
of QED was in  \cite{fock,klein,london,weyl,pauli}
(see also \cite{chern}).}.
The gauge principle allows very little arbitrariness: the interaction of matter
fields,  which carry non-commuting internal charges, and the nonlinear
self-interaction of gauge bosons are essentially fixed by the requirement
of local gauge invariance, very similar to the self-interaction of
gravitons in general relativity.

It is therefore appealing to extend the gauge principle, which was elevated by Yang and
Mills to a powerful constructive principle, so that it will define the
interaction of matter fields which carry
not only non-commutative internal charges, but
also arbitrary large spins. It seems that this will naturally
lead to a theory in which fundamental forces will be mediated by
integer-spin gauge quanta  and
that the Yang-Mills vector gauge boson will become a member of a bigger
family of tensor gauge bosons.

In the previous paper \cite{Savvidy:2005zm} we extended the
gauge principle so that it enlarges
the original algebra of the Abelian local gauge transformations
found in \cite{Savvidy:dv,Savvidy:2003fx,Savvidy:2005fe}
to a non-Abelian case. The extended non-Abelian gauge
transformations of the tensor gauge fields form a new large group.
On this large group one can define field strength tensors,
which are transforming homogeneously with respect to the extended
gauge transformations.  The invariant Lagrangian is quadratic in
the field strength tensors and describes interaction of tensor gauge
fields of arbitrary large integer spin $1,2,...$.

The purpose of the present paper is to present a second invariant Lagrangian
which can be constructed in terms of the above field strength tensors.
The total Lagrangian is a sum of the two Lagrangians and
exhibits enhanced local gauge invariance with double number of gauge parameters.
This allows to eliminate all negative norm states of the nonsymmetric second rank
tensor gauge field $A_{\mu\lambda} \neq A_{\lambda\mu}$, which describes therefore
three physical polarizations:  {\it two symmetric polarizations of
helicity-two massless charged tensor gauge bosons and
antisymmetric polarization of helicity-zero charged B field}.

The early investigation of higher-spin representations of the Poincar\'e
algebra and of the corresponding field equations is due to Majorana,
Dirac and Wigner \cite{majorana,dirac,wigner}. The theory of massive particles
of higher spin was further developed by Fierz and Pauli \cite{fierzpauli} and
Rarita and Schwinger \cite{rarita}. The Lagrangian and
S-matrix formulations of {\it free field
theory} of massive and massless fields with higher spin
have been completely constructed in
\cite{yukawa1,schwinger,Weinberg:1964cn,Weinberg:1964ev,Weinberg:1964ew,
chang,singh,singh1,fronsdal,fronsdal1}.
The problem of {\it introducing interaction} appears to be much more complex
\cite{Gupta,kraichnan,thirring,feynman,deser,fronsdal2,Sezgin:2001zs,Sagnotti:2005ns}
and met enormous difficulties for spin fields higher than two
\cite{witten,deser1,berends,dewit,vasiliev}.
The first positive result in this direction was
the light-front construction of the cubic
interaction term for the massless field of helicity $\pm \lambda$ in
\cite{Bengtsson:1983pd,Bengtsson:1983pg}.

In our approach the gauge fields are defined as rank-$(s+1)$ tensors
$$
A^{a}_{\mu\lambda_1 ... \lambda_{s}}(x),~~~~~s=0,1,2,...
$$
and are totally symmetric with respect to the
indices $  \lambda_1 ... \lambda_{s}  $. {\it A priory} the tensor fields
have no symmetries with
respect to the first index  $\mu$. This is an essential departure from the
previous considerations, in which the higher-rank tensors were totally symmetric
\cite{fierzpauli,schwinger,singh,fronsdal}.
The index $s$ runs from zero to infinity.
The first member of this family of the tensor gauge bosons is the Yang-Mills
vector boson $A^{a}_{\mu}$.

{\it The extended non-Abelian gauge transformations of the tensor gauge fields are defined
by the following equations } \cite{Savvidy:2005zm}:
\beqa\label{polygauge}
\delta A^{a}_{\mu} &=& ( \delta^{ab}\partial_{\mu}
+g f^{acb}A^{c}_{\mu})\xi^b ,~~~~~\\
\delta A^{a}_{\mu\nu} &=&  ( \delta^{ab}\partial_{\mu}
+  g f^{acb}A^{c}_{\mu})\xi^{b}_{\nu} + g f^{acb}A^{c}_{\mu\nu}\xi^{b},\nonumber\\
\delta A^{a}_{\mu\nu \lambda}& =&  ( \delta^{ab}\partial_{\mu}
+g f^{acb} A^{c}_{\mu})\xi^{b}_{\nu\lambda} +
g f^{acb}(  A^{c}_{\mu  \nu}\xi^{b}_{\lambda } +
A^{c}_{\mu \lambda }\xi^{b}_{ \nu}+
A^{c}_{\mu\nu\lambda}\xi^{b}),\nn\\
.........&.&............................\nn
\eeqa
or in the general form by the formula
\be\label{generalgaugetransform}
\delta A^{a}_{\mu\lambda_1 ... \lambda_s} = ( \delta^{ab}\partial_{\mu}
+g f^{acb} A^{c}_{\mu})\xi^{b}_{\lambda_1\lambda_2 ...\lambda_s} +
g f^{acb}~\sum^{s}_{i=1}  \sum_{P's} A^{c}_{\mu\lambda_1 ...\lambda_i}
\xi^{b}_{\lambda_{i+1} ...\lambda_s },
\ee
where the infinitesimal gauge parameters $\xi^{b}_{\lambda_{1} ...\lambda_s }$ are
totally symmetric rank-s tensors. The summation  $\sum_{P's}$ is
over all permutations of two
sets of indices $\lambda_1 ... \lambda_i$ and $\lambda_{i+1} ... \lambda_{s}$
which correspond to nonequal terms. It is obvious that this transformation
preserves the symmetry properties of the tensor gauge field
$A^{a}_{\mu\lambda_1 ... \lambda_s}$. Indeed, the first term in the r.h.s. is a covariant
derivative of the totally symmetric rank-s tensor
$\nabla^{ab}_{\mu}\xi^{b}_{\lambda_1\lambda_2 ...\lambda_s}$ and every term
$\sum_{P's} A^{c}_{\mu\lambda_1 ...\lambda_i}
\xi^{b}_{\lambda_{i+1} ...\lambda_s }$ in the second sum is totally
symmetric with respect to the indices $\lambda_1\lambda_2 ...\lambda_s$ by construction.
The matrix form of the transformation is
\beqa\label{matrixformofgaugetransformation}
\delta_{\xi} A_{\mu\lambda_1 ... \lambda_s} &=& \partial_{\mu}
\xi_{\lambda_1\lambda_2 ...\lambda_s}
-i g  [A_{\mu}, \xi_{\lambda_1\lambda_2 ...\lambda_s}] -i g
\sum^{s}_{i=1}  \sum_{P's} [A_{\mu\lambda_1 ...\lambda_i},
\xi_{\lambda_{i+1} ...\lambda_s }],
\eeqa
where the tensor gauge fields are
$A^{ab}_{\mu\lambda_1 ... \lambda_{s}} =
(L_c)^{ab}  A^{c}_{\mu\lambda_1 ... \lambda_{s}} = i f^{acb}A^{c}_{\mu
\lambda_1 ... \lambda_{s}}$,
and  $L^a$ are the generators of the semisimple Lie group G in the adjoint representation.

These extended gauge transformations
generate a closed algebraic structure. To see that, one should compute the
commutator of two extended gauge transformations $\delta_{\eta}$ and $\delta_{\xi}$
of parameters $\eta$ and $\xi$.
The commutator of two transformations can be expressed in the form \cite{Savvidy:2005zm}
\be\label{gaugecommutator}
[~\delta_{\eta},\delta_{\xi}]~A_{\mu\lambda_1\lambda_2 ...\lambda_s} ~=~
-i g~ \delta_{\zeta} A_{\mu\lambda_1\lambda_2 ...\lambda_s}
\ee
and is again an extended gauge transformation with the gauge parameters
$\{\zeta\}$ which are given by the matrix commutators
\beqa\label{gaugealgebra}
\zeta&=&[\eta,\xi]\\
\zeta_{\lambda_1}&=&[\eta,\xi_{\lambda_1}] +[\eta_{\lambda_1},\xi]\nn\\
\zeta_{\nu\lambda} &=& [\eta,\xi_{\nu\lambda}] +  [\eta_{\nu},\xi_{\lambda}]
+ [\eta_{\lambda},\xi_{\nu}]+[\eta_{\nu\lambda},\xi],\nn\\
......&.&..........................\nn
\eeqa
{\it The generalized field strengths  are defined as} \cite{Savvidy:2005zm}
\beqa\label{fieldstrengthparticular}
G^{a}_{\mu\nu} &=&
\partial_{\mu} A^{a}_{\nu} - \partial_{\nu} A^{a}_{\mu} +
g f^{abc}~A^{b}_{\mu}~A^{c}_{\nu},\\
G^{a}_{\mu\nu,\lambda} &=&
\partial_{\mu} A^{a}_{\nu\lambda} - \partial_{\nu} A^{a}_{\mu\lambda} +
g f^{abc}(~A^{b}_{\mu}~A^{c}_{\nu\lambda} + A^{b}_{\mu\lambda}~A^{c}_{\nu} ~),\nn\\
G^{a}_{\mu\nu,\lambda\rho} &=&
\partial_{\mu} A^{a}_{\nu\lambda\rho} - \partial_{\nu} A^{a}_{\mu\lambda\rho} +
g f^{abc}(~A^{b}_{\mu}~A^{c}_{\nu\lambda\rho} +
 A^{b}_{\mu\lambda}~A^{c}_{\nu\rho}+A^{b}_{\mu\rho}~A^{c}_{\nu\lambda}
 + A^{b}_{\mu\lambda\rho}~A^{c}_{\nu} ~),\nn\\
 ......&.&............................................\nn
\eeqa
and transform homogeneously with respect to the extended
gauge transformations (\ref{polygauge}). The field strength tensors are
antisymmetric in their first two indices and are totaly symmetric with respect to the
rest of the indices. By induction the entire construction
can be generalized to include tensor fields of any rank s, and
the corresponding field strength we shall define by the following expression:
\be\label{fieldstrengthgeneral}
G^{a}_{\mu\nu ,\lambda_1 ... \lambda_{s}} =
\partial_{\mu} A^{a}_{\nu \lambda_1 ... \lambda_{s}} -
\partial_{\nu} A^{a}_{\mu \lambda_1 ... \lambda_s} +
g f^{abc}\sum^{s}_{i=0}~ \sum_{P's} ~A^{b}_{\mu \lambda_1 ... \lambda_i}~
A^{c}_{\nu \lambda_{i+1} ... \lambda_{s}},
\ee
where the sum $\sum_{P's}$ runs over all permutations of two
sets of indices $\lambda_1 ... \lambda_i$ and $\lambda_{i+1} ... \lambda_{s}$
which correspond to nonequal terms.
All permutations of indices within two sets $\lambda_1 ... \lambda_i$ and
$\lambda_{i+1} ... \lambda_{s}$ correspond to equal terms, because
gauge fields are totally symmetric with respect to $\lambda_1 ... \lambda_i$ and
$\lambda_{i+1} ... \lambda_{s}$. Therefore there are
$$
\frac{s!}{i!(s-i)!}
$$
nonequal terms in the sum $\sum_{P's}$. Thus in the sum $\sum_{P's}$ there is one
term of the form $A_{\mu}A_{\nu\lambda_1\lambda_{2} ... \lambda_{s}}$, there are
$s$ terms
of the form $A_{\mu \lambda_1}
A_{\nu \lambda_{2} ... \lambda_{s}}$ and $s(s-1)/2$ terms of the form
$A_{\mu \lambda_1 \lambda_2}~
A_{\nu \lambda_{3} ... \lambda_{s}}$ and so on.
In the above definition of the extended gauge field strength
$G^{a}_{\mu\nu ,\lambda_1 ... \lambda_{s}}$,
together with the classical Yang-Mills gauge boson
$A^{a}_{\mu}$, there participate a set of higher-rank gauge fields
$A^{a}_{\mu\lambda_1}$, $A^{a}_{\mu\lambda_1 , \lambda_2}$,  ... ,
$A^{a}_{\mu\lambda_1 ...\lambda_{s}}$ up to the rank $s+1$.
By construction the field strength (\ref{fieldstrengthgeneral})
is antisymmetric with respect to its first two indices
$
G^{a}_{\mu\nu ,\lambda_1 ... \lambda_{s}}~ = ~-~G^{a}_{\nu \mu,\lambda_1 ... \lambda_{s}}
$
and is totally symmetric with respect to the rest of the indices
$
G^{a}_{\mu\nu ,..\lambda_{i}...\lambda_{j}.. } =
G^{a}_{\mu\nu ,..\lambda_{j}...\lambda_{i}..}~,
$
where $1 \leq i < j \leq s$.

The inhomogeneous extended gauge transformation (\ref{generalgaugetransform})
induces the homogeneous gauge
transformation of the corresponding field strength
(\ref{fieldstrengthgeneral}) of the form \cite{Savvidy:2005zm}
\beqa\label{fieldstrenghparticular}
\delta G^{a}_{\mu\nu}&=& g f^{abc} G^{b}_{\mu\nu} \xi^c \\
\delta G^{a}_{\mu\nu,\lambda} &=& g f^{abc} (~G^{b}_{\mu\nu,\lambda} \xi^c
+ G^{b}_{\mu\nu} \xi^{c}_{\lambda}~),\nonumber\\
\delta G^{a}_{\mu\nu,\lambda\rho} &=& g f^{abc}
(~G^{b}_{\mu\nu,\lambda\rho} \xi^c
+ G^{b}_{\mu\nu,\lambda} \xi^{c}_{\rho} +
G^{b}_{\mu\nu,\rho} \xi^{c}_{\lambda} +
G^{b}_{\mu\nu} \xi^{c}_{\lambda\rho}~)\nn\\
......&.&..........................,\nn
\eeqa
or in general
\be\label{variationfieldstrengthgeneral}
\delta G^{a}_{\mu\nu,\lambda_1 ... \lambda_s} =
g f^{abc} \sum^{s}_{i=0}~ \sum_{P's} ~G^{b}_{\mu\nu,\lambda_1 ... \lambda_i}
\xi^{c}_{\lambda_{i+1}...\lambda_s}.
\ee
The symmetry properties of the field strength  $G^{a}_{\mu\nu,\lambda_1 ... \lambda_s}$
remain invariant in the course of this transformation.
The polygauge invariant Lagrangian  now can be formulated in the form \cite{Savvidy:2005zm}
\beqa\label{fulllagrangian1}
{{\cal L}}_{s+1}&=&-{1\over 4} ~
G^{a}_{\mu\nu, \lambda_1 ... \lambda_s}~
G^{a}_{\mu\nu, \lambda_{1}...\lambda_{s}} +.......\nonumber\\
&=& -{1\over 4}\sum^{2s}_{i=0}~a^{s}_i ~
G^{a}_{\mu\nu, \lambda_1 ... \lambda_i}~
G^{a}_{\mu\nu, \lambda_{i+1}...\lambda_{2s}}
\sum_{p's} \eta^{\lambda_{i_1} \lambda_{i_2}} .......
\eta^{\lambda_{i_{2s-1}} \lambda_{i_{2s}}}~,
\eeqa
where the sum $\sum_p$ runs over all permutations of indices
$\lambda_1 ... \lambda_{2s}$ which correspond to nonequal terms. For the low
values of $s=0,1,2,...$ the numerical coefficients $$a^{s}_i = {s!\over i!(2s-i)!}$$
are: $a^{0}_0=1;~~a^{1}_1 =1,a^{1}_0 =a^{1}_2 =1/2;~~
a^{2}_2 =1/2,a^{2}_1 =a^{2}_3 =1/3,a^{2}_0 =a^{2}_4 =1/12;$ and so on.
In order to describe fixed rank-$(s+1)$ gauge field
one should have  at disposal all gauge fields
up to the rank $2s+1$.
In order to make all tensor gauge fields dynamical one should add
the corresponding kinetic terms. Thus the invariant
Lagrangian describing dynamical tensor gauge bosons of all ranks
has the form
\be\label{fulllagrangian2}
{{\cal L}} = \sum^{\infty}_{s=1}~ g_s {{\cal L}}_s~
\ee
where ${{\cal L}}_1$ is the Yang-Mills Lagrangian, ${{\cal L}}_2$,
${{\cal L}}_3$,... are new invariant forms quadratic in field strength tensors
(\ref{fulllagrangian1}) and $g_s$ are new coupling constants ($g_1 =1$).

  The first three terms of the invariant Lagrangian have
the following form \cite{Savvidy:2005zm}:
\beqa\label{firstthreeterms}
{{\cal L}} =  {{\cal L}}_1 + g_2 {{\cal L}}_2 +g_3 {{\cal L}}_3 +...
=-{1\over 4}G^{a}_{\mu\nu} G^{a}_{\mu\nu}
+g_2\{-{1\over 4}G^{a}_{\mu\nu,\lambda}G^{a}_{\mu\nu,\lambda}
-{1\over 4}G^{a}_{\mu\nu}G^{a}_{\mu\nu,\lambda\lambda}~\}~~~~~~~~~~\nn\\
+g_3\{-{1\over 4}G^{a}_{\mu\nu,\lambda\rho}G^{a}_{\mu\nu,\lambda\rho}
-{1\over 8}G^{a}_{\mu\nu ,\lambda\lambda}G^{a}_{\mu\nu ,\rho\rho}
-{1\over 2}G^{a}_{\mu\nu,\lambda}  G^{a}_{\mu\nu ,\lambda \rho\rho}
-{1\over 8}G^{a}_{\mu\nu}  G^{a}_{\mu\nu ,\lambda \lambda\rho\rho}~\}+..,
\eeqa
where the first term is the Yang-Mills Lagrangian and the
second and the third ones describe the tensor gauge fields $A^{a}_{\mu\nu},
A^{a}_{\mu\nu\lambda}$ and so on.
It is important that:  i) {\it the Lagrangian does not
contain higher derivatives of tensor gauge fields
ii) all interactions take place
through the three- and four-particle exchanges with dimensionless
coupling constant g  iii) the complete Lagrangian contains all higher-rank
tensor gauge fields and should not be truncated iv) $g_s$ are new
coupling constants}.

\section{{\it Enhanced Local Gauge Symmetry}}

The Lagrangian (\ref{fulllagrangian1}), (\ref{fulllagrangian2})
and (\ref{firstthreeterms}) is not the
most general Lagrangian which can be constructed in terms
of the above field strength tensors
(\ref{fieldstrengthparticular}) and (\ref{fieldstrengthgeneral}). We shall
see that there exists a second invariant Lagrangian
${{\cal L}}^{'}_{2}$ which can be constructed for the second-rank
nonsymmetric tensor gauge field
$A^{a}_{\mu\lambda} \neq A^{a}_{\lambda\mu}$ and that the total Lagrangian is a
sum of two Lagrangians ${{\cal L}}_{2}+{{\cal L}}^{'}_{2} $ and
exhibits a new enhanced gauge symmetry (\ref{largegaugetransformation})
which allows to eliminate all negative norm
polarizations of the nonsymmetric second-rank tensor gauge field $A_{\mu\lambda}$.

The invariance of the Lagrangian
$$
{{\cal L}}_2 =-{1\over 4}G^{a}_{\mu\nu,\lambda}G^{a}_{\mu\nu,\lambda}
-{1\over 4}G^{a}_{\mu\nu}G^{a}_{\mu\nu,\lambda\lambda}
$$
in (\ref{fulllagrangian1}), (\ref{fulllagrangian2}) and (\ref{firstthreeterms})
was demonstrated in \cite{Savvidy:2005zm} by
calculating its variation with respect to the gauge
transformation (\ref{polygauge}) and (\ref{fieldstrenghparticular}),
(\ref{variationfieldstrengthgeneral}).
Indeed, its variation is equal to zero
\beqa
\delta {{\cal L}}_2 =
&-&{1\over 2} G^{a}_{\mu\nu,\lambda} g f^{abc} (G^{b}_{\mu\nu,\lambda} \xi^c  +
G^{b}_{\mu\nu} \xi^{c}_{\lambda})
-{1\over 4} g f^{abc} G^{b}_{\mu\nu} \xi^c G^{a}_{\mu\nu,\lambda\lambda}\nonumber\\
&-&{1\over 4} G^{a}_{\mu\nu} g f^{abc} (G^{b}_{\mu\nu,\lambda\lambda} \xi^c  +
2 G^{b}_{\mu\nu, \lambda} \xi^{c}_{\lambda}+
G^{b}_{\mu\nu} \xi^{c}_{\lambda \lambda})=0 .\nonumber
\eeqa
But the Lagrangian ${{\cal L}}_{2}$
is not a unique one, there exist additional Lorentz invariant
quadratic forms which can be constructed by the corresponding field
strength tensors. They are
$$
G^{a}_{\mu\nu,\lambda}G^{a}_{\mu\lambda,\nu},~~~
G^{a}_{\mu\nu,\nu}G^{a}_{\mu\lambda,\lambda},~~~
G^{a}_{\mu\nu}G^{a}_{\mu\lambda,\nu\lambda}.
$$
Calculating  the variation of each of these terms with respect to
the gauge transformation (\ref{polygauge}) and (\ref{fieldstrenghparticular})
one can get convinced that a particular linear combination
\be\label{actiontwoprime}
{{\cal L}}^{'}_2 =  {1\over 4}G^{a}_{\mu\nu,\lambda}G^{a}_{\mu\lambda,\nu}
+{1\over 4}G^{a}_{\mu\nu,\nu}G^{a}_{\mu\lambda,\lambda}
+{1\over 2} G^{a}_{\mu\nu}G^{a}_{\mu\lambda,\nu\lambda}
\ee
forms an invariant Lagrangian. Indeed, the variation of the Lagrangian ${{\cal L}}^{'}_2$
under the gauge transformation (\ref{fieldstrenghparticular}) is equal to zero
\beqa
\delta {{\cal L}}^{'}_2 =
&+&{1\over 4} G^{a}_{\mu\nu,\lambda} g f^{abc} (G^{b}_{\mu\lambda,\nu} \xi^c  +
G^{b}_{\mu\lambda} \xi^{c}_{\nu})
+{1\over 4} g f^{abc} (G^{b}_{\mu\nu,\lambda} \xi^c  +
G^{b}_{\mu\nu} \xi^{c}_{\lambda}) G^{a}_{\mu\lambda,\nu}    \nn\\
&+& {1\over 2}G^{a}_{\mu\nu,\nu}g f^{abc} (G^{b}_{\mu\lambda,\lambda}  \xi^c  +
G^{b}_{\mu\lambda} \xi^{c}_{\lambda})\nn\\
&+& {1\over 2}g f^{abc} G^{b}_{\mu\nu} \xi^c G^{a}_{\mu\lambda,\nu\lambda}\nonumber\\
&+& {1\over 2} G^{a}_{\mu\nu} g f^{abc} (G^{b}_{\mu\lambda,\nu\lambda} \xi^c  +
G^{b}_{\mu\lambda,\nu } \xi^{c}_{\lambda}+
G^{b}_{\mu\lambda,\lambda } \xi^{c}_{\nu}+
G^{b}_{\mu\lambda} \xi^{c}_{\nu \lambda})=0 .\nonumber
\eeqa
As a result we have two invariant Lagrangians
${{\cal L}}_2$ and ${{\cal L}}^{'}_2$ and the general Lagrangian is a
linear combination of these two invariant forms
$
g_2 {{\cal L}}_2 + g^{'}_{2} {{\cal L}}^{'}_2 ,
$
where $g_2$ and $g^{'}_{2}$  are arbitrary constants.

{\it Our aim now is to demonstrate  that
if $g_2 = g^{'}_{2}$ then we shall have enhanced local gauge invariance
(\ref{largegaugetransformation}) of the Lagrangian
${{\cal L}}_2 + {{\cal L}}^{'}_2$ with double number of gauge parameters.
This allows to eliminate all negative norm states of the nonsymmetric second-rank
tensor gauge field $A^{a}_{\mu \lambda} \neq A^{a}_{\lambda\mu}$, which describes
therefore three physical modes: two symmetric polarizations of
helicity-two massless charged tensor gauge bosons and
antisymmetric polarization of helicity-zero charged B field.}

Indeed let us consider the
situation at the linearized level when the gauge coupling constant g is equal to zero.
The free part of the ${{\cal L}}_2$ Lagrangian
$$
{{\cal L}}^{free}_2 ={1 \over 2} A^{a}_{\alpha\acute{\alpha}}
(\eta_{\alpha\gamma}\eta_{\acute{\alpha}\acute{\gamma}}\partial^{2} -
\eta_{\acute{\alpha}\acute{\gamma}} \partial_{\alpha} \partial_{\gamma} )
A^{a}_{\gamma\acute{\gamma}} =
{1 \over 2} A^{a}_{\alpha\acute{\alpha}}
H_{\alpha\acute{\alpha}\gamma\acute{\gamma}} A^{a}_{\gamma\acute{\gamma}} ,
$$
where the quadratic form in the momentum representation has the form
$$
H_{\alpha\acute{\alpha}\gamma\acute{\gamma}}(k)=
(-k^2 \eta_{\alpha\gamma} +k_{\alpha}k_{\gamma})
\eta_{\acute{\alpha}\acute{\gamma}},
$$
is obviously invariant with respect to the gauge
transformation $\delta A^{a}_{\mu\lambda} =\partial_{\mu} \xi^{a}_{\lambda}$,
but it is not invariant with respect to the alternative gauge transformations
$\delta A^{a}_{\mu \lambda} =\partial_{\lambda} \eta^{a}_{\mu}$. This can be
seen, for example, from the following relations in momentum representation:
\be\label{currentdivergence}
k_{\alpha}H_{\alpha\acute{\alpha}\gamma\acute{\gamma}}(k)=0,~~~
k_{\acute{\alpha}}H_{\alpha\acute{\alpha}\gamma\acute{\gamma}}(k)=
-(k^2 \eta_{\alpha\gamma} - k_{\alpha}k_{\gamma})k_{\acute{\gamma}} \neq 0 .
\ee
Let us consider now the free part of the second Lagrangian
\beqa
{{\cal L}}^{' free}_{2} ={1 \over 4} A^{a}_{\alpha\acute{\alpha}}
(-\eta_{\alpha\acute{\gamma}}\eta_{\acute{\alpha}\gamma}\partial^{2} -
\eta_{\alpha\acute{\alpha}}\eta_{\gamma\acute{\gamma}}\partial^{2}
+\eta_{\alpha\acute{\gamma}} \partial_{\acute{\alpha}} \partial_{\gamma}
+\eta_{\acute{\alpha}\gamma} \partial_{\alpha} \partial_{\acute{\gamma}}
+\eta_{\alpha\acute{\alpha}} \partial_{\gamma} \partial_{\acute{\gamma}}+\nn\\
+\eta_{\gamma\acute{\gamma}} \partial_{\alpha} \partial_{\acute{\alpha}}
-2\eta_{\alpha\gamma} \partial_{\acute{\alpha}} \partial_{\acute{\gamma}})
A^{a}_{\gamma\acute{\gamma}}=
{1 \over 2} A^{a}_{\alpha\acute{\alpha}}
H^{~'}_{\alpha\acute{\alpha}\gamma\acute{\gamma}} A^{a}_{\gamma\acute{\gamma}},
\eeqa
where
$$
H^{'}_{\alpha\acute{\alpha}\gamma\acute{\gamma}}(k)=
{1 \over 2}(\eta_{\alpha\acute{\gamma}}\eta_{\acute{\alpha}\gamma}
+\eta_{\alpha\acute{\alpha}}\eta_{\gamma\acute{\gamma}})k^2
-{1 \over 2}(\eta_{\alpha\acute{\gamma}}k_{\acute\alpha}k_{\gamma}
+\eta_{\acute\alpha\gamma}k_{\alpha}k_{\acute{\gamma}}
+\eta_{\alpha\acute\alpha}k_{\gamma}k_{\acute{\gamma}}
+\eta_{\gamma\acute{\gamma}}k_{\alpha}k_{\acute\alpha}
-2\eta_{\alpha\gamma}k_{\acute\alpha}k_{\acute{\gamma}}).
$$
It is again invariant with respect to the gauge
transformation $\delta A^{a}_{\mu\lambda} =\partial_{\mu} \xi^{a}_{\lambda}$,
but it is not invariant with respect to the gauge transformations
$\delta A^{a}_{\mu \lambda} =\partial_{\lambda} \eta^{a}_{\mu}$ as one can
see from analogous relations
\be\label{currentdivergenceprime}
k_{\alpha}H^{'}_{\alpha\acute{\alpha}\gamma\acute{\gamma}}(k)=0,~~~
k_{\acute{\alpha}}H^{'}_{\alpha\acute{\alpha}\gamma\acute{\gamma}}(k)=
(k^2 \eta_{\alpha\gamma} -k_{\alpha}k_{\gamma})k_{\acute{\gamma}} \neq 0 .
\ee
As it is obvious from (\ref{currentdivergence}) and
(\ref{currentdivergenceprime}), the total Lagrangian
${{\cal L}}^{free}_2 + {{\cal L}}^{' free}_2$ now  poses new enhanced
invariance with respect to the larger, eight parameter, gauge transformations
\be\label{largegaugetransformation}
\delta A^{a}_{\mu \lambda} =\partial_{\mu} \xi^{a}_{\lambda}+
\partial_{\lambda} \eta^{a}_{\mu} +...,
\ee
where $\xi^{a}_{\lambda}$ and $\eta^{a}_{\mu}$ are eight arbitrary functions, because
\be\label{zeroderivatives}
k_{\alpha}(H_{\alpha\acute{\alpha}\gamma\acute{\gamma}}+
H^{'}_{\alpha\acute{\alpha}\gamma\acute{\gamma}})=0,~~~
k_{\acute{\alpha}}(H_{\alpha\acute{\alpha}\gamma\acute{\gamma}}+
H^{'}_{\alpha\acute{\alpha}\gamma\acute{\gamma}})=0 .
\ee
Thus our free part of the Lagrangian is
\beqa\label{totalfreelagrangian}
{{\cal L}}^{tot~free}_{2} =&-&{1 \over 2}\partial_{\mu}
A^{a}_{\nu \lambda}\partial_{\mu} A^{a}_{\nu \lambda}
+{1 \over 2}\partial_{\mu} A^{a}_{\nu \lambda}\partial_{\nu} A^{a}_{\mu \lambda}+
\nn\\
&+&{1 \over 4} \partial_{\mu} A^{a}_{\nu \lambda} \partial_{\mu } A^{a}_{\lambda\nu}
-{1 \over 4} \partial_{\mu} A^{a}_{\nu \lambda} \partial_{\lambda} A^{a}_{\mu \nu}
-{1 \over 4}\partial_{\nu} A^{a}_{\mu \lambda} \partial_{\mu} A^{a}_{\lambda\nu }
+{1 \over 4} \partial_{\nu } A^{a}_{\mu\lambda} \partial_{\lambda} A^{a}_{\mu \nu}
\nn\\
&+&{1 \over 4}\partial_{\mu} A^{a}_{\nu \nu}\partial_{\mu} A^{a}_{\lambda\lambda}
-{1 \over 2}\partial_{\mu} A^{a}_{\nu \nu} \partial_{\lambda} A^{a}_{\mu\lambda}
+{1 \over 4}\partial_{\nu } A^{a}_{\mu\nu}\partial_{\lambda} A^{a}_{\mu\lambda}
\eeqa
or, in equivalent form, it is
\beqa\label{totfreelagrangianalternativeform}
{{\cal L}}^{tot~free}_{2} ={1 \over 2} A^{a}_{\alpha\acute{\alpha}}
\{(\eta_{\alpha\gamma}\eta_{\acute{\alpha}\acute{\gamma}}
-{1\over 2}\eta_{\alpha\acute{\gamma}}\eta_{\acute{\alpha}\gamma}
-{1\over 2}\eta_{\alpha\acute{\alpha}}\eta_{\gamma\acute{\gamma}})
\partial^{2}
-\eta_{\acute{\alpha}\acute{\gamma}} \partial_{\alpha} \partial_{\gamma}
-\eta_{\alpha\gamma} \partial_{\acute{\alpha}} \partial_{\acute{\gamma}}+\nn\\
+{1\over 2}(\eta_{\alpha\acute{\gamma}} \partial_{\acute{\alpha}} \partial_{\gamma}
+\eta_{\acute{\alpha}\gamma} \partial_{\alpha} \partial_{\acute{\gamma}}
+\eta_{\alpha\acute{\alpha}} \partial_{\gamma} \partial_{\acute{\gamma}}
+\eta_{\gamma\acute{\gamma}} \partial_{\alpha} \partial_{\acute{\alpha}})
\}
A^{a}_{\gamma\acute{\gamma}}
\eeqa
and is invariant with respect to the larger gauge transformations
$
\delta A^{a}_{\mu \lambda} =\partial_{\mu} \xi^{a}_{\lambda}+
\partial_{\lambda} \eta^{a}_{\mu},
$
where $\xi^{a}_{\lambda}$ and $\eta^{a}_{\mu}$ are eight arbitrary functions.
In momentum representation the quadratic form is
\beqa\label{quadraticform}
H^{tot}_{\alpha\acute{\alpha}\gamma\acute{\gamma}}(k)=
(-\eta_{\alpha\gamma}\eta_{\acute{\alpha}\acute{\gamma}}
+{1 \over 2}\eta_{\alpha\acute{\gamma}}\eta_{\acute{\alpha}\gamma}
+{1 \over 2}\eta_{\alpha\acute{\alpha}}\eta_{\gamma\acute{\gamma}})k^2
+\eta_{\alpha\gamma}k_{\acute\alpha}k_{\acute{\gamma}}
+\eta_{\acute\alpha \acute{\gamma}}k_{\alpha}k_{\gamma}\nn\\
-{1 \over 2}(\eta_{\alpha\acute{\gamma}}k_{\acute\alpha}k_{\gamma}
+\eta_{\acute\alpha\gamma}k_{\alpha}k_{\acute{\gamma}}
+\eta_{\alpha\acute\alpha}k_{\gamma}k_{\acute{\gamma}}
+\eta_{\gamma\acute{\gamma}}k_{\alpha}k_{\acute\alpha}).
\eeqa
In summary, we have the following Lagrangian for the
lower-rank tensor gauge fields:
\beqa\label{totalactiontwo}
{{\cal L}}=  {{\cal L}}_1 +  g_2({{\cal L}}_2 +  {{\cal L}}^{'}_2) &=&
-{1\over 4}G^{a}_{\mu\nu}G^{a}_{\mu\nu}\\
&+&g_2 \{-{1\over 4}G^{a}_{\mu\nu,\lambda}G^{a}_{\mu\nu,\lambda}
-{1\over 4}G^{a}_{\mu\nu}G^{a}_{\mu\nu,\lambda\lambda} +\nn\\
&&+{1\over 4}G^{a}_{\mu\nu,\lambda}G^{a}_{\mu\lambda,\nu}
+{1\over 4}G^{a}_{\mu\nu,\nu}G^{a}_{\mu\lambda,\lambda}
+{1\over 2}G^{a}_{\mu\nu}G^{a}_{\mu\lambda,\nu\lambda}~\}.\nn
\eeqa
Let us consider the equations of motion which follow from this Lagrangian for
the vector gauge field $A^{a}_{\nu}$
\beqa\label{equationforfirstranktensor}
&&\nabla^{ab}_{\mu}G^{b}_{\mu\nu}
+g_2 \{{1\over 2 }\nabla^{ab}_{\mu} (G^{b}_{\mu\nu,\lambda\lambda}
+ G^{b}_{\nu\lambda,\mu\lambda}
+ G^{b}_{\lambda\mu,\nu\lambda})
+ g f^{acb} A^{c}_{\mu\lambda} G^{b}_{\mu\nu,\lambda}\\
&-&{1\over 2 }g f^{acb} (A^{c}_{\mu\lambda} G^{b}_{\mu\lambda,\nu}
+A^{c}_{\mu\lambda} G^{b}_{\lambda\nu,\mu}
+A^{c}_{\lambda\lambda} G^{b}_{\mu\nu,\mu}
+A^{c}_{\mu\nu} G^{b}_{\mu\lambda,\lambda}-\nn\\
&&~~~~~~~~~~~-A^{c}_{\mu\lambda\lambda} G^{b}_{\mu\nu}
- A^{c}_{\mu\mu\lambda} G^{b}_{\nu\lambda}
- A^{c}_{\mu\nu\lambda} G^{b}_{\lambda\mu}) \}\nn
=0
\eeqa\label{equationforsecondranktensor}
and for the second-rank tensor gauge field $A^{a}_{\nu\lambda}$
\beqa
&&\nabla^{ab}_{\mu}G^{b}_{\mu\nu,\lambda}
-{1\over 2} (\nabla^{ab}_{\mu}G^{b}_{\mu\lambda,\nu}
+\nabla^{ab}_{\mu}G^{b}_{\lambda\nu,\mu}
+\nabla^{ab}_{\lambda}G^{b}_{\mu\nu,\mu}
+\eta_{\nu\lambda} \nabla^{ab}_{\mu}G^{b}_{\mu\rho,\rho})\nn\\
&+&g f^{acb} A^{c}_{\mu\lambda} G^{b}_{\mu\nu} -
{1\over 2}g f^{acb}(A^{c}_{\mu\nu} G^{b}_{\mu\lambda}
+A^{c}_{\mu\mu} G^{b}_{\lambda\nu}
+A^{c}_{\lambda\mu} G^{b}_{\mu\nu}
+\eta_{\nu\lambda}  A^{c}_{\mu\rho} G^{b}_{\mu\rho})
=0.
\eeqa
Representing these equations in the form
\beqa\label{freeequations}
\partial_{\mu} F^{a}_{\mu\nu}
+g_2 {1\over 2 }\partial_{\mu} (F^{a}_{\mu\nu,\lambda\lambda}
+ F^{a}_{\nu\lambda,\mu\lambda}
+ F^{a}_{\lambda\mu,\nu\lambda}) = j^{a}_{\nu},~~~~\nn\\
\partial_{\mu} F^{a}_{\mu\nu,\lambda}
-{1\over 2} (\partial_{\mu} F^{a}_{\mu\lambda,\nu}
+\partial_{\mu} F^{a}_{\lambda\nu,\mu}
+\partial_{\lambda}F^{a}_{\mu\nu,\mu}
+\eta_{\nu\lambda} \partial_{\mu}F^{a}_{\mu\rho,\rho}) = j^{a}_{\nu\lambda},
\eeqa
where $F^{a}_{\mu\nu} =  \partial_{\mu} A^{a}_{\nu  } -
\partial_{\nu} A^{a}_{\mu },~
F^{a}_{\mu\nu,\lambda} = \partial_{\mu} A^{a}_{\nu \lambda} -
\partial_{\nu} A^{a}_{\mu \lambda},~
F^{a}_{\mu\nu,\lambda\rho} = \partial_{\mu} A^{a}_{\nu \lambda\rho} -
\partial_{\nu} A^{a}_{\mu \lambda\rho}$ ,
we can find the corresponding conserved currents
\beqa
j^{a}_{\nu } = &-&g f^{abc} A^{b}_{\mu } G^{c}_{\mu\nu }
-g f^{abc}\partial_{\mu} (A^{b}_{\mu } A^{c}_{\nu })\nn\\
&-&{1\over 2 }g f^{abc}A^{b}_{\mu} (G^{c}_{\mu\nu,\lambda\lambda}
+ G^{c}_{\nu\lambda,\mu\lambda}
+ G^{c}_{\lambda\mu,\nu\lambda}
)
-{1\over 2 }\partial_{\mu} (I^{a}_{\mu\nu,\lambda\lambda}
+ I^{a}_{\nu\lambda,\mu\lambda}
+ I^{a}_{\lambda\mu,\nu\lambda})\nn\\
&-& g f^{abc} A^{b}_{\mu\lambda} G^{c}_{\mu\nu,\lambda}
+ {1\over 2 }g f^{abc} (A^{b}_{\mu\lambda} G^{c}_{\mu\lambda,\nu}
+A^{b}_{\mu\lambda} G^{c}_{\lambda\nu,\mu}
+A^{b}_{\lambda\lambda} G^{c}_{\mu\nu,\mu}
+A^{b}_{\mu\nu} G^{c}_{\mu\lambda,\lambda}
) \nn\\
&-&{1\over 2} g f^{abc} (A^{b}_{\mu\lambda\lambda} G^{c}_{\mu\nu}
+ A^{b}_{\lambda\mu\lambda} G^{c}_{\nu\mu}
+ A^{b}_{\mu\lambda\nu} G^{c}_{\lambda\mu}),
\eeqa
where $I^{a}_{\mu\nu,\lambda\rho}=g f^{abc}(~A^{b}_{\mu}~A^{c}_{\nu\lambda\rho} +
 A^{b}_{\mu\lambda}~A^{c}_{\nu\rho}+A^{b}_{\mu\rho}~A^{c}_{\nu\lambda}
 + A^{b}_{\mu\lambda\rho}~A^{c}_{\nu} ~)$ and
\beqa
j^{a}_{\nu\lambda}=&-&g f^{abc} A^{b}_{\mu} G^{c}_{\mu\nu,\lambda}
+{1\over 2 }g f^{abc} (A^{b}_{\mu} G^{c}_{\mu\lambda,\nu}
+A^{b}_{\mu} G^{c}_{\lambda\nu,\mu}
+A^{b}_{\lambda} G^{c}_{\mu\nu,\mu}
+\eta_{\nu\lambda}A^{b}_{\mu} G^{c}_{\mu\rho,\rho})\nn\\
&-&g f^{abc} A^{b}_{\mu\lambda} G^{c}_{\mu\nu} +
{1\over 2}g f^{abc}(A^{b}_{\mu\nu} G^{c}_{\mu\lambda}
+A^{b}_{\lambda\mu} G^{c}_{\mu\nu}
+A^{b}_{\mu\mu} G^{c}_{\lambda\nu}
+\eta_{\nu\lambda}  A^{b}_{\mu\rho} G^{c}_{\mu\rho})\nn\\
&-&g f^{abc} \partial_{\mu}
(A^{b}_{\mu} A^{c}_{\nu\lambda} + A^{b}_{\mu\lambda} A^{c}_{\nu}) +
{1\over 2}g f^{abc}
[\partial_{\mu}(A^{b}_{\mu} A^{c}_{\lambda\nu}+A^{b}_{\mu\nu} A^{c}_{\lambda})
+\partial_{\mu}(A^{b}_{\lambda} A^{c}_{\nu\mu}+A^{b}_{\lambda\mu} A^{c}_{\nu})\nn\\
&+&\partial_{\lambda} (A^{b}_{\mu} A^{c}_{\nu\mu}  + A^{b}_{\mu\mu} A^{c}_{\nu})
+\eta_{\nu\lambda}  \partial_{\mu}
(A^{b}_{\mu} A^{c}_{\rho\rho} + A^{b}_{\mu\rho} A^{c}_{\rho})].
\eeqa
Thus on mass shell
$$
\partial_{\nu} j^{a}_{\nu}=0,~~~\partial_{\nu} j^{a}_{\nu\lambda}=0,~~~~
\partial_{\lambda} j^{a}_{\nu\lambda}=0,
$$
because, as we demonstrated above, the partial derivatives of the l.h.s.'s of the equations
(\ref{freeequations}) are equal to zero (see equations (\ref{zeroderivatives})).

At the linearized level, when the gauge coupling constant g is equal to zero,
the equations of motion (\ref{freeequations}) for the second-rank tensor
gauge fields will take the form
\beqa\label{mainequation}
\partial^{2}(A^{a}_{\nu\lambda} -{1\over 2}A^{a}_{\lambda\nu})
-\partial_{\nu} \partial_{\mu}  (A^{a}_{\mu\lambda}-
{1\over 2}A^{a}_{\lambda\mu} )&-&
\partial_{\lambda} \partial_{\mu}  (A^{a}_{\nu\mu} - {1\over 2}A^{a}_{\mu\nu} )
+\partial_{\nu} \partial_{\lambda} ( A^{a}_{\mu\mu}-{1\over 2}A^{a}_{\mu\mu})\nn\\
&+&{1\over 2}\eta_{\nu\lambda} ( \partial_{\mu} \partial_{\rho}A^{a}_{\mu\rho}
-  \partial^{2}A^{a}_{\mu\mu})=0
\eeqa
and, as we shall see below, describe the propagation of massless particles
of spin 2 and spin 0. Indeed, it is easy to see that our equation
reduces to the well known Fierz-Pauli-Schwinger-Chang-Singh-Hagen-Fronsdal equation
for the symmetric part of the tensor gauge fields
$A^{a}_{\nu\lambda}$
$$
\partial^{2} A_{\nu\lambda}
-\partial_{\nu} \partial_{\mu}  A_{\mu\lambda} -
\partial_{\lambda} \partial_{\mu}  A_{\mu\nu}
+ \partial_{\nu} \partial_{\lambda}  A_{\mu\mu}
+\eta_{\nu\lambda}  (\partial_{\mu} \partial_{\rho}A_{\mu\rho}
- \partial^{2} A_{\mu\mu}) =0,
$$
which describes the propagation of massless boson with two
physical polarizations, the $\lambda= \pm 2$ helicity states.
For the antisymmetric part $A^{a}_{\nu\lambda}$
the equation reduces to the form
\be\label{antisymmetric}
\partial^{2} A_{\nu\lambda}
-\partial_{\nu} \partial_{\mu}  A_{\mu\lambda} +
\partial_{\lambda} \partial_{\mu}  A_{\mu\nu}=0
\ee
and describes the propagation of massless tensor boson with
one physical polarization, the $\lambda= 0$ helicity state.
Thus we have seen that the extended gauge symmetry is sufficient to
exclude all negative norm polarizations from the spectrum of the second-rank
{\it nonsymmetric tensor gauge field} $A_{\mu\lambda}$.

Using the gauge invariance (\ref{largegaugetransformation}) we can impose the
Lorentz invariant supplementary
conditions on the second-rank gauge fields $A_{\mu\lambda}$:
$
\partial_{\mu} A^{a}_{\mu\lambda} =a_{\lambda} ,~~
\partial_{\lambda} A^{a}_{\mu\lambda} =b_{\mu} ,
$
where $a_{\lambda}$ and $b_{\mu}$ are arbitrary functions, or
one can suggest alternative
supplementary conditions in which the quadratic form
(\ref{totalfreelagrangian}), (\ref{totfreelagrangianalternativeform}),
(\ref{quadraticform}) is diagonal:
\be\label{diagonalgauge}
\partial_{\mu} A^{a}_{\mu\lambda} -{1\over 2} \partial_{\lambda} A^{a}_{\mu\mu}=0,~~
\partial_{\lambda} A^{a}_{\mu\lambda} -{1\over 2} \partial_{\mu} A^{a}_{\lambda\lambda}=0.
\ee
In this gauge the equation  has the  form
\beqa\label{gaugefixedequations}
\partial^{2} A^{a}_{\nu\lambda}  =0
\eeqa
and in the momentum representation
$
A_{\mu\nu}(x) = e_{\mu\nu}(k) e^{ikx}
$
from  equation (\ref{gaugefixedequations}) it follows that $k^2=0$ and
we have {\it massless particles}.
In this gauge we shall get
$$
H^{fix}_{\alpha\acute{\alpha}\gamma\acute{\gamma}}(k) =
(\eta_{\alpha\gamma}\eta_{\acute{\alpha}\acute{\gamma}}
-{1\over 2}\eta_{\alpha\acute{\alpha}}\eta_{\gamma\acute{\gamma}})(-k^2 )
$$
and can find the propagator $\Delta_{\gamma\acute{\gamma}\lambda\acute{\lambda}}(k)$
from the equation
$
H^{fix}_{\alpha\acute{\alpha}\gamma\acute{\gamma}}(k)
\Delta_{\gamma\acute{\gamma}\lambda\acute{\lambda}}(k) =
\eta_{\alpha\lambda}\eta_{\acute{\alpha}\acute{\lambda}}~~,
$
thus
\be
\Delta_{\gamma\acute{\gamma}\lambda\acute{\lambda}}(k) = -
{\eta_{\gamma\lambda}\eta_{\acute{\gamma}\acute{\lambda}}
-{1\over 2}\eta_{\gamma\acute{\gamma}}\eta_{\lambda\acute{\lambda}}
 \over k^2 - i\varepsilon}~~.
\ee
The corresponding residue can be represented as a sum
\beqa
\eta_{\gamma\lambda}\eta_{\acute{\gamma}\acute{\lambda}}
-{1\over 2}\eta_{\gamma\acute{\gamma}}\eta_{\lambda\acute{\lambda}}
=&+&{1\over 2}(\eta_{\gamma\lambda}\eta_{\acute{\gamma}\acute{\lambda}}
+\eta_{\gamma\acute{\lambda}}\eta_{\acute{\gamma}\lambda}
-\eta_{\gamma\acute{\gamma}}\eta_{\lambda\acute{\lambda}})+
 {1\over 2}(\eta_{\gamma\lambda}\eta_{\acute{\gamma}\acute{\lambda}}
-\eta_{\gamma\acute{\lambda}}\eta_{\acute{\gamma}\lambda}).
\eeqa
The first term describes the $\lambda= \pm 2$ helicity states and is
represented by the symmetric part of the polarization tensor $e_{\mu\lambda}$,
the second term describes $\lambda= 0$ helicity state and is represented
by its antisymmetric part.
When $k_{\mu}$ is aligned along the third axis,
$k_{\mu}= (k,0,0,k)$, we have two independent polarizations of the helicity-2
and helicity-0 particles
\beqa
e^{1}_{\mu\lambda}={1\over \sqrt{2}}
\left( \begin{array}{llll}
  0,0,~~0,0\\
  0,1,~~0,0\\
  0,0,-1,0\\
  0,0,~~0,0
\end{array} \right), e^{2}_{\mu\lambda}={1\over \sqrt{2}}
\left( \begin{array}{ll}
  0,0,0,0\\
  0,0,1,0\\
  0,1,0,0\\
  0,0,0,0
\end{array} \right),
e^{A}_{\mu\lambda}={1\over \sqrt{2}}
\left( \begin{array}{ll}
  0,~~0,0,0\\
  0,~~0,1,0\\
  0,-1,0,0\\
  0,~~0,0,0
\end{array} \right),\nn\\
\eeqa
with the property that
$
e^{1}_{\gamma\acute{\gamma}}e^{1}_{\lambda\acute{\lambda}}  +
e^{2}_{\gamma\acute{\gamma}}e^{2}_{\lambda\acute{\lambda}} \simeq
{1\over 2}(\eta_{\gamma\lambda}\eta_{\acute{\gamma}\acute{\lambda}}
+\eta_{\gamma\acute{\lambda}}\eta_{\acute{\gamma}\lambda}
-\eta_{\gamma\acute{\gamma}}\eta_{\lambda\acute{\lambda}})
$
and
$
e^{A}_{\gamma\acute{\gamma}}e^{A}_{\lambda\acute{\lambda}}
\simeq{1\over 2}(\eta_{\gamma\lambda}\eta_{\acute{\gamma}\acute{\lambda}}
-\eta_{\gamma\acute{\lambda}}\eta_{\acute{\gamma}\lambda}).
$
The symbol $\simeq$ means that the equation holds up to longitudinal terms.

Thus the general second-rank tensor gauge field with 16
components $A_{\mu\lambda}$ describes in this theory three
physical propagating massless polarizations.

\section{{\it Interaction Vertices}}
We are interested now to analyze the interaction properties
of the tensor gauge bosons prescribed by the gauge principle.
The interaction of the Yang-Mills vector bosons with
the charged tensor gauge bosons is described by the Lagrangian
(\ref{totalactiontwo})
$
{{\cal L}}_2 +  {{\cal L}}^{'}_2 .
$
Let us first consider a three-particle interaction vertices - VTT.
Explicitly three-linear terms of the Lagrangian ${{\cal L}}_2$
have the form:
\beqa\label{cubicterm}
{{\cal L}}^{cubic}_{2}  =
&-&{1 \over 2}g f^{abc}(\partial_{\mu} A^{a}_{\nu\lambda} -
\partial_{\nu} A^{a}_{\mu\lambda})~ (A^{b}_{\mu}A^{c}_{\nu\lambda}+
A^{b}_{\mu\lambda}A^{c}_{\nu})\nonumber\\
&-&{1 \over 4}g f^{abc}(\partial_{\mu} A^{a}_{\nu} -
\partial_{\nu} A^{a}_{\mu})~ 2A^{b}_{\mu\lambda}A^{c}_{\nu\lambda},
\eeqa
and in addition to the standard Yang-Mills VVV three vector boson interaction vertex
$$
{{\cal L}}^{cubic}_{1} = -{1 \over 2} g f^{abc}(\partial_{\mu} A^{a}_{\nu} -
\partial_{\nu} A^{a}_{\mu})
A^{b}_{\mu} A^{c}_{\nu},
$$
which in momentum representation has the form
\be
{{\cal V}}^{abc}_{\alpha\beta\gamma}(k,p,q) =
-i g f^{abc} [\eta_{\alpha\beta} (p-k)_{\gamma}+ \eta_{\alpha\gamma} (k-q)_{\beta}
 + \eta_{\beta\gamma} (q-p)_{\alpha}] = -i g f^{abc}
 F_{\alpha\beta\gamma}(k,p,q),
\ee
we have a new three-particle interaction vertex of one vector boson and
two tensor gauge bosons - VTT. In momentum space it has the form
\be
{{\cal V}}^{abc}_{\alpha\acute{\alpha}\beta\gamma\acute{\gamma}}(k,p,q) =
- i g f^{abc} [\eta_{\alpha\beta} (p-k)_{\gamma}+ \eta_{\alpha\gamma} (k-q)_{\beta}
 + \eta_{\beta\gamma} (q-p)_{\alpha}] \eta_{\acute{\alpha}\acute{\gamma}}.
\ee
Notice that two parts in (\ref{cubicterm}), which came from  different
terms of the Lagrangian ${{\cal L}}_2$, combine into the VVV vertex and the tensor
$\eta_{\acute{\alpha}\acute{\gamma}}$. It is convenient to represent the
vertex in the form
\be
{{\cal V}}^{abc}_{\alpha\acute{\alpha}\beta\gamma\acute{\gamma}}(k,p,q) =
-i g f^{abc} F_{\alpha\acute{\alpha}\beta\gamma\acute{\gamma}}(k,p,q).
\ee
We have also a three-particle interaction vertex VTT of one vector boson and
two tensor gauge bosons in the second Lagrangian ${{\cal L}}^{'}_2$. Explicitly the
three-linear terms of Lagrangian ${{\cal L}}^{'}_2$ have the form:

\beqa
{{\cal L}}^{'~cubic}_{2} &=&{1\over 2}g f^{abc}(\partial_{\mu} A^{a}_{\nu\lambda} -
\partial_{\nu} A^{a}_{\mu\lambda})~ (A^{b}_{\mu}A^{c}_{\lambda\nu}+
A^{b}_{\mu\nu}A^{c}_{\lambda})\nonumber\\
&+&{1\over 2}g f^{abc}(\partial_{\mu} A^{a}_{\nu\nu} -
\partial_{\nu} A^{a}_{\mu\nu})~ (A^{b}_{\mu}A^{c}_{\lambda\lambda}+
A^{b}_{\mu\lambda}A^{c}_{\lambda})\nonumber\\
&+&{1\over 2}g f^{abc}(\partial_{\mu} A^{a}_{\nu} -
\partial_{\nu} A^{a}_{\mu})~ (A^{b}_{\mu\nu}A^{c}_{\lambda\lambda}+
A^{b}_{\mu\lambda}A^{c}_{\lambda\nu}) ,
\eeqa
so that in momentum space we have
\beqa
{{\cal V}}^{'abc}_{\alpha\acute{\alpha}\beta\gamma\acute{\gamma}}(k,p,q)
 &=&{1\over 2} i g f^{abc}
F^{'}_{\alpha\acute{\alpha}\beta\gamma\acute{\gamma}}(k,p,q)\nn\\
F^{'}_{\alpha\acute{\alpha}\beta\gamma\acute{\gamma}}(k,p,q) &=&
(p-k)_{\gamma}(\eta_{\alpha\acute{\gamma}}
\eta_{\acute{\alpha}\beta}+
\eta_{\alpha\acute{\alpha}} \eta_{\beta\acute{\gamma}})\nn\\
&+& (k-q)_{\beta}(\eta_{\alpha\acute{\gamma}} \eta_{\acute{\alpha}\gamma}+
\eta_{\alpha\acute{\alpha}} \eta_{\gamma\acute{\gamma}})\nn\\
&+& (q-p)_{\alpha} (\eta_{\acute{\alpha}\gamma} \eta_{\beta\acute{\gamma}}+
\eta_{\acute{\alpha}\beta} \eta_{\gamma\acute{\gamma}})\nn\\
&+&(p-k)_{\acute{\alpha}}\eta_{\alpha\beta} \eta_{\gamma\acute{\gamma}}+
(p-k)_{\acute{\gamma}} \eta_{\alpha\beta} \eta_{\acute{\alpha}\gamma}\nn\\
&+&(k-q)_{\acute{\alpha}} \eta_{\alpha\gamma} \eta_{\beta\acute{\gamma}}+
(k-q)_{\acute{\gamma}}\eta_{\alpha\gamma} \eta_{\acute{\alpha}\beta}\nn\\
&+&(q-p)_{\acute{\alpha}} \eta_{\beta\gamma} \eta_{\alpha\acute{\gamma}}+
(q-p)_{\acute{\gamma}}\eta_{\alpha\acute{\alpha}} \eta_{\beta\gamma}.
\eeqa
Collecting two terms of the three-point vertex VTT together we shall get
\be
{{\cal V}}^{tot~abc}_{\alpha\acute{\alpha}\beta\gamma\acute{\gamma}}(k,p,q) =
{{\cal V}}^{abc}_{\alpha\acute{\alpha}\beta\gamma\acute{\gamma}}(k,p,q) +
{{\cal V}}^{'abc}_{\alpha\acute{\alpha}\beta\gamma\acute{\gamma}}(k,p,q).
\ee

Let us consider now  four-particle interaction terms of the Lagrangian
${{\cal L}}_2 +  {{\cal L}}^{'}_2$. We have the
standard  four vector boson interaction vertex VVVV
\beqa
{{\cal V}}^{abcd}_{\alpha\beta\gamma\delta}(k,p,q,r) = - g^2 f^{lac}f^{lbd} (\eta_{\alpha \beta}
\eta_{\gamma\delta} - \eta_{\alpha \delta} \eta_{\beta\gamma})\nonumber\\
-g^2 f^{lad}f^{lbc} (\eta_{\alpha \beta} \eta_{\gamma\delta} -
\eta_{\alpha \gamma}\eta_{\beta\delta} )\nonumber\\
-g^2 f^{lab}f^{lcd} (\eta_{\alpha \gamma} \eta_{\beta\delta} -
\eta_{\alpha \delta}\eta_{\beta\gamma} )
\eeqa
and a new interaction of two vector and two tensor gauge bosons - the VVTT vertex,
\beqa
{{\cal L}}^{quartic}_{2} =
&-&{1 \over 4}g^2 f^{abc}f^{a\acute{b}\acute{c}}
(A^{b}_{\mu} A^{c}_{\nu\lambda} +
A^{b}_{\mu\lambda} A^{c}_{\nu})(
A^{\acute{b}}_{\mu}A^{\acute{c}}_{\nu\lambda} +
~A^{\acute{b}}_{\mu\lambda}A^{\acute{c}}_{\nu})\nonumber\\
&-&{1 \over 2}g^2 f^{abc}f^{a\acute{b}\acute{c}}
A^{b}_{\mu} A^{c}_{\nu}A^{\acute{b}}_{\mu\lambda}A^{\acute{c}}_{\nu\lambda},
\eeqa
which in the momentum space will take the form
\beqa
{{\cal V}}^{abcd}_{\alpha\beta\gamma\acute{\gamma}\delta\acute{\delta}}(k,p,q,r)=
&-&  g^2 f^{lac}f^{lbd} (\eta_{\alpha \beta}
\eta_{\gamma\delta} - \eta_{\alpha \delta} \eta_{\beta\gamma})
\eta_{\acute{\gamma}\acute{\delta}}\nonumber\\
&-& g^2 f^{lad}f^{lbc} (\eta_{\alpha \beta} \eta_{\gamma\delta} -
\eta_{\alpha \gamma}\eta_{\beta\delta} )\eta_{\acute{\gamma}\acute{\delta}}\nonumber\\
&-& g^2 f^{lab}f^{lcd} (\eta_{\alpha \gamma} \eta_{\beta\delta} -
\eta_{\alpha \delta}\eta_{\beta\gamma} )\eta_{\acute{\gamma}\acute{\delta}}.
\eeqa
The second part of the vertex VVTT comes from the Lagrangian ${{\cal L}}^{'}_2$:
\beqa
{{\cal L}}^{' quartic}_2=&+&{1 \over 4}g^2 f^{abc}f^{a\acute{b}\acute{c}}
(A^{b}_{\mu} A^{c}_{\nu\lambda} +
A^{b}_{\mu\lambda} A^{c}_{\nu})
(A^{\acute{b}}_{\mu}A^{\acute{c}}_{\lambda\nu} +
~A^{\acute{b}}_{\mu\nu}A^{\acute{c}}_{\lambda})\nonumber\\
&+&{1 \over 4}g^2 f^{abc}f^{a\acute{b}\acute{c}}
(A^{b}_{\mu} A^{c}_{\nu\nu} + A^{b}_{\mu\nu} A^{c}_{\nu})
(A^{\acute{b}}_{\mu}A^{\acute{c}}_{\lambda\lambda} +
~A^{\acute{b}}_{\mu\lambda}A^{\acute{c}}_{\lambda})\nonumber\\
&+&{1 \over 2}g^2 f^{abc}f^{a\acute{b}\acute{c}}
A^{b}_{\mu} A^{c}_{\nu}
(A^{\acute{b}}_{\mu\nu}A^{\acute{c}}_{\lambda\lambda} +
A^{\acute{b}}_{\mu\lambda}A^{\acute{c}}_{\lambda\nu}),
\eeqa
which in the momentum representation will take the form
\beqa
2 {{\cal V}}^{'~abcd}_{\alpha\beta\gamma\acute{\gamma}\delta\acute{\delta}}(k,p,q,r)=
g^2 f^{lac}f^{lbd} [&+&\eta_{\alpha \beta}
(\eta_{\gamma\acute{\delta}}\eta_{\acute{\gamma}\delta} +
\eta_{\gamma\acute{\gamma}} \eta_{\delta\acute{\delta}})\nn\\
&-&\eta_{\beta\gamma}
(\eta_{\alpha \acute{\delta}}\eta_{\acute{\gamma}\delta} +
\eta_{\alpha\acute{\gamma}} \eta_{\delta\acute{\delta}})\nn\\
&-&\eta_{\alpha\delta}
(\eta_{\beta\acute{\gamma}}\eta_{\gamma\acute{\delta}} +
\eta_{\beta\acute{\delta}} \eta_{\gamma\acute{\gamma}})\nn\\
&+&\eta_{\gamma\delta}
(\eta_{\alpha\acute{\delta}}\eta_{\beta\acute{\gamma}} +
\eta_{\alpha\acute{\gamma}} \eta_{\beta\acute{\delta}})]\nn\\
g^2 f^{lad}f^{lbc} [&+&\eta_{\alpha \beta}
(\eta_{\gamma\acute{\delta}}\eta_{\acute{\gamma}\delta} +
\eta_{\gamma\acute{\gamma}} \eta_{\delta\acute{\delta}})\nn\\
&-&\eta_{\alpha\gamma}
(\eta_{\beta\acute{\delta}}\eta_{\acute{\gamma}\delta} +
\eta_{\beta\acute{\gamma}} \eta_{\delta\acute{\delta}})\nn\\
&-&\eta_{\beta\delta}
(\eta_{\alpha\acute{\gamma}}\eta_{\gamma\acute{\delta}} +
\eta_{\alpha\acute{\delta}} \eta_{\gamma\acute{\gamma}})\nn\\
&+&\eta_{\gamma\delta}
(\eta_{\alpha\acute{\gamma}}\eta_{\beta\acute{\delta}} +
\eta_{\alpha\acute{\delta}} \eta_{\beta\acute{\gamma}})]\nn\\
g^2 f^{lab}f^{lcd}[&+&\eta_{\alpha \gamma}
(\eta_{\beta\acute{\gamma}}\eta_{\delta\acute{\delta}} +
\eta_{\beta\acute{\delta}} \eta_{\delta\acute{\gamma}})\nn\\
&-&\eta_{\beta\gamma}
(\eta_{\alpha\acute{\gamma}}\eta_{\delta\acute{\delta}} +
\eta_{\alpha\acute{\delta}} \eta_{\delta\acute{\gamma}})\nn\\
&-&\eta_{\alpha\delta}
(\eta_{\beta\acute{\delta}}\eta_{\gamma\acute{\gamma}} +
\eta_{\beta\acute{\gamma}} \eta_{\gamma\acute{\delta}})\nn\\
&+&\eta_{\beta\delta}
(\eta_{\alpha\acute{\delta}}\eta_{\gamma\acute{\gamma}} +
\eta_{\alpha\acute{\gamma}} \eta_{\gamma\acute{\delta}})].
\eeqa
The total vertex is
\be
{{\cal V}}^{tot~abcd}_{\alpha\beta\gamma\delta}(k,p,q,r)=
{{\cal V}}^{abcd}_{\alpha\beta\gamma\delta}(k,p,q,r) +
{{\cal V}}^{'~abcd}_{\alpha\beta\gamma\delta}(k,p,q,r).
\ee

\section{{\it Third-Rank Tensor Gauge Fields}}
The Lagrangian $ {{\cal L}}_1 +  g_2({{\cal L}}_2 +  {{\cal L}}^{'}_2)$ contains the
third-rank gauge fields $A^{a}_{\mu\nu\lambda}$, but without the corresponding
kinetic term. In order to make the fields $A^{a}_{\mu\nu\lambda}$ dynamical
we have added the corresponding Lagrangian ${{\cal L}}_3$ presented at the second line
of the formula (\ref{firstthreeterms}).
But again the Lagrangian
${{\cal L}}_3$ is not the most general invariant
which can be constructed from the corresponding field strength tensors. There are
seven Lorentz invariant quadratic forms which form the second invariant
Lagrangian ${{\cal L}}^{'}_3$ so that at this level the total Lagrangian is a sum
$$
{{\cal L}}= {{\cal L}}_1 +  g_2({{\cal L}}_2 +
{{\cal L}}^{'}_2) +g_3 {{\cal L}}_3+ g^{'}_{3}{{\cal L}}^{'}_3 +...
$$

Indeed, the Lagrangian ${{\cal L}}_3$ has the form (\ref{firstthreeterms}):
\beqa
{{\cal L}}_3 =-{1\over 4}G^{a}_{\mu\nu,\lambda\rho}G^{a}_{\mu\nu,\lambda\rho}
-{1\over 8}G^{a}_{\mu\nu ,\lambda\lambda}G^{a}_{\mu\nu ,\rho\rho}
-{1\over 2}G^{a}_{\mu\nu,\lambda}  G^{a}_{\mu\nu ,\lambda \rho\rho}
-{1\over 8}G^{a}_{\mu\nu}  G^{a}_{\mu\nu ,\lambda \lambda\rho\rho}~,
\eeqa
where the field strength tensors (\ref{fieldstrengthgeneral}) are
\beqa\label{spin4fieldstrenghth}
G^{a}_{\mu\nu ,\lambda \rho \sigma} =
\partial_{\mu} A^{a}_{\nu \lambda \rho \sigma} -
\partial_{\nu} A^{a}_{\mu \lambda\rho\sigma} +
g f^{abc}\{~A^{b}_{\mu}~A^{c}_{\nu \lambda \rho\sigma}
+A^{b}_{\mu\lambda}~A^{c}_{\nu\rho \sigma} +
A^{b}_{\mu\rho }~A^{c}_{\nu\lambda\sigma} +
A^{b}_{\mu\sigma}~A^{c}_{\nu\lambda\rho} +\nn\\
+A^{b}_{\mu\lambda\rho}~A^{c}_{\nu \sigma} +
A^{b}_{\mu\lambda\sigma}~A^{c}_{\nu\rho} +
A^{b}_{\mu\rho\sigma}~A^{c}_{\nu \lambda} +
     A^{b}_{\mu\lambda\rho\sigma }~A^{c}_{\nu} ~\}\nonumber
\eeqa
and
\beqa\label{spin4fieldstrenghth4}
G^{a}_{\mu\nu ,\lambda \rho \sigma\delta} =
\partial_{\mu} A^{a}_{\nu \lambda \rho \sigma\delta} -
\partial_{\nu} A^{a}_{\mu \lambda\rho\sigma\delta} &+&
g f^{abc}\{~A^{b}_{\mu}~A^{c}_{\nu \lambda \rho\sigma\delta}
+\sum_{ \lambda \leftrightarrow \rho,\sigma,\delta}
        A^{b}_{\mu\lambda}~A^{c}_{\nu\rho \sigma\delta} + \nn\\
&+&\sum_{\lambda,\rho \leftrightarrow \sigma,\delta}
        A^{b}_{\mu\lambda\rho}~A^{c}_{\nu\sigma\delta} +
\sum_{\lambda,\rho,\sigma\leftrightarrow \delta}
       A^{b}_{\mu\lambda\rho\sigma}~A^{c}_{\nu\delta} +
     A^{b}_{\mu\lambda\rho\sigma\delta }~A^{c}_{\nu} ~\}.\nonumber
\eeqa
The terms in parenthesis are symmetric over $\lambda \rho\sigma$ and
$\lambda \rho \sigma\delta$ respectively. The Lagrangian ${{\cal L}}_3$
is invariant with respect to the extended gauge transformations  (\ref{polygauge})
of the low-rank gauge fields
$ A_{\mu}, A_{\mu\nu}, A_{\mu\nu\lambda}$  and of the fourth-rank gauge field
(\ref{matrixformofgaugetransformation})
\beqa\label{gaugetransform4}
\delta_{\xi}  A_{\mu\nu\lambda\rho} =\partial_{\mu}\xi_{\nu\lambda\rho}
-i g[A_{\mu},\xi_{\nu\lambda\rho}]
-i g [A_{\mu\nu},\xi_{\lambda\rho}]
-i g [A_{\mu\lambda},\xi_{\nu\rho}]
-i g [A_{\mu\rho},\xi_{\nu\lambda}]-\nn\\
-i g  [A_{\mu\nu\lambda},\xi_{\rho}]
-i g  [A_{\mu\nu\rho},\xi_{\lambda}]
-i g  [A_{\mu\lambda\rho},\xi_{\nu}]
-i g [A_{\mu\nu\lambda\rho},\xi]\nonumber
\eeqa
and also of the fifth-rank tensor gauge field (\ref{matrixformofgaugetransformation})
\beqa\label{gaugetransform5}
\delta_{\xi}  A_{\mu\nu\lambda\rho\sigma} &=&\partial_{\mu}\xi_{\nu\lambda\rho\sigma}
-i g[A_{\mu},\xi_{\nu\lambda\rho\sigma}]
-i g \sum_{\nu \leftrightarrow \lambda\rho\sigma}
[A_{\mu\nu},\xi_{\lambda\rho\sigma}]-\nn\\
   &~&-ig\sum_{\nu\lambda \leftrightarrow \rho\sigma}
   [A_{\mu\nu\lambda},\xi_{\rho\sigma}]
      -ig\sum_{\nu\lambda\rho \leftrightarrow \sigma}
      [A_{\mu\nu\lambda\rho},\xi_{\sigma}]
-i g [A_{\mu\nu\lambda\rho},\xi], ~\nonumber
\eeqa
where the gauge parameters $\xi_{\nu\lambda\rho}$ and $\xi_{\nu\lambda\rho\sigma}$
are totally symmetric rank-3 and rank-4 tensors.
The extended gauge transformation of the higher-rank tensor gauge
fields induces the gauge transformation of the fields strengths of the form
(\ref{variationfieldstrengthgeneral})
\beqa\label{spin4fieldstrenghthtransfor}
\delta G^{a}_{\mu\nu,\lambda\rho\sigma} =
g f^{abc} (~G^{b}_{\mu\nu,\lambda\rho\sigma} ~\xi^c  +
 G^{b}_{\mu\nu, \lambda\rho} ~\xi^{c}_{\sigma}+
G^{b}_{\mu\nu, \lambda\sigma} ~\xi^{c}_{\rho}+
G^{b}_{\mu\nu, \rho\sigma} ~\xi^{c}_{\lambda}+~~~~~~~~~~~~~~~~~~~~~~~~~\nn\\
+ G^{b}_{\mu\nu,\lambda } ~\xi^{c}_{\rho\sigma}+
 G^{b}_{\mu\nu,\rho} ~\xi^{c}_{\lambda\sigma}+
 G^{b}_{\mu\nu,\sigma} ~\xi^{c}_{\lambda \rho}+
G^{b}_{\mu\nu } ~\xi^{c}_{\lambda\rho\sigma}~)\nonumber
\eeqa
and
\beqa\label{fieldstrengh5thtransfor}
\delta G^{a}_{\mu\nu,\lambda\rho\sigma\delta} =
g f^{abc} (~G^{b}_{\mu\nu,\lambda\rho\sigma\delta} ~\xi^c
&+& \sum_{ \lambda\rho,\sigma \leftrightarrow \delta}
G^{b}_{\mu\nu, \lambda\rho\sigma} ~\xi^{c}_{\delta}
+ \nn\\
&+& \sum_{ \lambda\rho \leftrightarrow \sigma,\delta}
G^{b}_{\mu\nu, \lambda\rho} ~\xi^{c}_{\sigma\delta}+
         \sum_{ \lambda \leftrightarrow \rho,\sigma,\delta}
         G^{b}_{\mu\nu, \lambda} ~\xi^{c}_{\rho\sigma\delta}~+
         G^{b}_{\mu\nu } ~\xi^{c}_{\lambda\rho\sigma\delta}).\nonumber
\eeqa
Using the above homogeneous transformations for the field strength
tensors one can demonstrate the invariance of the
Lagrangian ${{\cal L}}_3$ with respect to the extended gauge transformations
(\ref{fieldstrenghparticular}),
(\ref{spin4fieldstrenghthtransfor}) and  (\ref{fieldstrengh5thtransfor})
(see reference \cite{Savvidy:2005zm} for details).

Our purpose now is to present a second invariant Lagrangian
which can be constructed in terms of the above field strength tensors.
Let us consider the following seven Lorentz invariant quadratic forms which
can be constructed by the corresponding field strength tensors
\beqa
G^{a}_{\mu\nu,\lambda\rho}G^{a}_{\mu\lambda,\nu\rho},~~~
G^{a}_{\mu\nu,\nu\lambda}G^{a}_{\mu\rho,\rho\lambda},~~~
G^{a}_{\mu\nu,\nu\lambda}G^{a}_{\mu\lambda,\rho\rho},~~~
G^{a}_{\mu\nu,\lambda}G^{a}_{\mu\lambda,\nu\rho\rho},~~~\nn\\
G^{a}_{\mu\nu,\lambda}G^{a}_{\mu\rho,\nu\lambda\rho},~~~
G^{a}_{\mu\nu,\nu}G^{a}_{\mu\lambda,\lambda\rho\rho},~~~
G^{a}_{\mu\nu}G^{a}_{\mu\lambda,\nu\lambda\rho\rho}.~~~~~~~~~~~~~~~~~
\eeqa
Calculating  the variation of each of these terms with respect to
the gauge transformation (\ref{fieldstrenghparticular}),
(\ref{spin4fieldstrenghthtransfor}) and  (\ref{fieldstrengh5thtransfor})
one can get convinced that the particular linear combination
\beqa\label{actionthreeprime}
{{\cal L}}^{'}_3 &=&  {1\over 4}
G^{a}_{\mu\nu,\lambda\rho}G^{a}_{\mu\lambda,\nu\rho}+
{1\over 4} G^{a}_{\mu\nu,\nu\lambda}G^{a}_{\mu\rho,\rho\lambda}+
{1\over 4}G^{a}_{\mu\nu,\nu\lambda}G^{a}_{\mu\lambda,\rho\rho}\nn\\
&+&{1\over 4}G^{a}_{\mu\nu,\lambda}G^{a}_{\mu\lambda,\nu\rho\rho}
+{1\over 2}G^{a}_{\mu\nu,\lambda}G^{a}_{\mu\rho,\nu\lambda\rho}
+{1\over 4}G^{a}_{\mu\nu,\nu}G^{a}_{\mu\lambda,\lambda\rho\rho}
+{1\over 4}G^{a}_{\mu\nu}G^{a}_{\mu\lambda,\nu\lambda\rho\rho}
\eeqa
forms an invariant Lagrangian.  Indeed, the variation of the first term is
$$
\delta_{\xi} G^{a}_{\mu\nu,\lambda\rho}G^{a}_{\mu\lambda,\nu\rho}=
2g f^{abc}G^{a}_{\mu\nu,\lambda\rho}G^{b}_{\mu\lambda,\nu}\xi^{c}_{\rho}+
2g f^{abc}G^{a}_{\mu\nu,\lambda\rho}G^{b}_{\mu\lambda,\rho}\xi^{c}_{\nu}+
2g f^{abc}G^{a}_{\mu\nu,\lambda\rho}G^{b}_{\mu\lambda}\xi^{c}_{\nu\rho} ,
$$
of the second term is
$$
\delta_{\xi} G^{a}_{\mu\nu,\nu\lambda}G^{a}_{\mu\rho,\rho\lambda}=
2g f^{abc}G^{a}_{\mu\nu,\nu\lambda}G^{b}_{\mu\rho,\rho}\xi^{c}_{\lambda}+
2g f^{abc}G^{a}_{\mu\nu,\nu\lambda}G^{b}_{\mu\rho,\lambda}\xi^{c}_{\rho}+
2g f^{abc}G^{a}_{\mu\nu,\nu\lambda}G^{b}_{\mu\rho}\xi^{c}_{\rho\lambda} ,
$$
of the third term is
\beqa
\delta_{\xi} G^{a}_{\mu\nu,\nu\lambda}G^{a}_{\mu\lambda,\rho\rho}=
2g f^{abc}G^{a}_{\mu\nu,\nu\lambda}G^{b}_{\mu\lambda,\rho}\xi^{c}_{\rho}+
g f^{abc}G^{a}_{\mu\nu,\nu\lambda}G^{b}_{\mu\lambda}\xi^{c}_{\rho\rho}+
g f^{abc}G^{a}_{\mu\lambda,\rho\rho}G^{b}_{\mu\nu,\nu}\xi^{c}_{\lambda}+\nn\\
+g f^{abc}G^{a}_{\mu\lambda,\rho\rho}G^{b}_{\mu\nu,\lambda}\xi^{c}_{\nu}+
g f^{abc}G^{a}_{\mu\lambda,\rho\rho}G^{b}_{\mu\nu}\xi^{c}_{\nu\lambda}\nn ,
\eeqa
of the forth term is
\beqa
\delta_{\xi} G^{a}_{\mu\nu,\lambda}G^{a}_{\mu\lambda,\nu\rho\rho}=
g f^{abc}G^{a}_{\mu\lambda,\nu\rho\rho}G^{b}_{\mu\nu}\xi^{c}_{\lambda}+
2g f^{abc}G^{a}_{\mu\lambda,\nu\rho}G^{b}_{\mu\nu,\lambda}\xi^{c}_{\rho}+
g f^{abc}G^{a}_{\mu\lambda,\rho\rho}G^{b}_{\mu\nu,\lambda}\xi^{c}_{\nu}+\nn\\
+g f^{abc}G^{a}_{\mu\nu,\lambda}G^{b}_{\mu\lambda,\nu}\xi^{c}_{\rho\rho}+
2g f^{abc}G^{a}_{\mu\nu,\lambda}G^{b}_{\mu\lambda,\rho}\xi^{c}_{\nu\rho}+
g f^{abc}G^{a}_{\mu\nu,\lambda}G^{b}_{\mu\lambda}\xi^{c}_{\nu\rho\rho}\nn ,
\eeqa
of the fifth term is
\beqa
\delta_{\xi} G^{a}_{\mu\nu,\lambda}G^{a}_{\mu\rho,\nu\lambda\rho}=\nn\\
g f^{abc}G^{a}_{\mu\rho,\nu\lambda\rho}G^{b}_{\mu\nu}\xi^{c}_{\lambda}+
g f^{abc}G^{b}_{\mu\rho,\nu\lambda}G^{a}_{\mu\nu,\lambda}\xi^{c}_{\rho}+
g f^{abc}G^{b}_{\mu\rho,\nu\rho}G^{a}_{\mu\nu,\lambda}\xi^{c}_{\lambda}+
g f^{abc}G^{b}_{\mu\rho,\lambda\rho}G^{a}_{\mu\nu,\lambda}\xi^{c}_{\nu}+\nn\\
+g f^{abc}G^{a}_{\mu\nu,\lambda}G^{b}_{\mu\rho,\nu}\xi^{c}_{\lambda\rho}+
g f^{abc}G^{a}_{\mu\nu,\lambda}G^{b}_{\mu\rho,\lambda}\xi^{c}_{\nu\rho}+
g f^{abc}G^{a}_{\mu\nu,\lambda}G^{b}_{\mu\rho,\rho}\xi^{c}_{\nu\lambda}+
g f^{abc}G^{a}_{\mu\nu,\lambda}G^{b}_{\mu\rho}\xi^{c}_{\nu\lambda\rho}\nn ,
\eeqa
of the sixth term is
\beqa
\delta_{\xi} G^{a}_{\mu\nu,\nu}G^{a}_{\mu\lambda,\lambda\rho\rho}=\nn\\
g f^{abc}G^{a}_{\mu\lambda,\lambda\rho\rho}G^{b}_{\mu\nu}\xi^{c}_{\nu}+
2g f^{abc}G^{b}_{\mu\lambda,\lambda\rho}G^{a}_{\mu\nu,\nu}\xi^{c}_{\rho}+
g f^{abc}G^{b}_{\mu\lambda,\rho\rho}G^{a}_{\mu\nu,\nu}\xi^{c}_{\lambda}+
g f^{abc}G^{b}_{\mu\lambda,\lambda}G^{a}_{\mu\nu,\nu}\xi^{c}_{\rho\rho}+\nn\\
+2g f^{abc}G^{b}_{\mu\lambda,\rho}G^{a}_{\mu\nu,\nu}\xi^{c}_{\lambda\rho}+
g f^{abc}G^{a}_{\mu\nu,\nu}G^{b}_{\mu\lambda}\xi^{c}_{\lambda\rho\rho}\nn
\eeqa
and finally of the seventh term is
\beqa
\delta_{\xi} G^{a}_{\mu,\nu}G^{a}_{\mu\lambda,\nu\lambda\rho\rho}=\nn\\
2g f^{abc}G^{a}_{\mu\nu}G^{b}_{\mu\lambda,\nu\lambda\rho}\xi^{c}_{\rho}+
g f^{abc}G^{a}_{\mu\nu}G^{b}_{\mu\lambda,\nu\rho\rho}\xi^{c}_{\lambda}+
g f^{abc}G^{a}_{\mu\nu}G^{b}_{\mu\lambda,\lambda\rho\rho}\xi^{c}_{\nu}+
g f^{abc}G^{a}_{\mu\nu}G^{b}_{\mu\lambda,\nu\lambda}\xi^{c}_{\rho\rho}+\nn\\
2g f^{abc}G^{a}_{\mu\nu}G^{b}_{\mu\lambda,\nu\rho}\xi^{c}_{\lambda\rho}+
2g f^{abc}G^{a}_{\mu\nu}G^{b}_{\mu\lambda,\lambda\rho}\xi^{c}_{\nu\rho}+
g f^{abc}G^{a}_{\mu\nu}G^{b}_{\mu\lambda,\rho\rho}\xi^{c}_{\nu\lambda}+\nn\\
g f^{abc}G^{a}_{\mu\nu}G^{b}_{\mu\lambda,\nu}\xi^{c}_{\lambda\rho\rho}+
g f^{abc}G^{a}_{\mu\nu}G^{b}_{\mu\lambda,\lambda}\xi^{c}_{\nu\rho\rho}
+2g f^{abc}G^{a}_{\mu\nu}G^{b}_{\mu\lambda,\rho}\xi^{c}_{\nu\lambda\rho}+
g f^{abc}G^{a}_{\mu\nu}G^{b}_{\mu\lambda}\xi^{c}_{\nu\lambda\rho\rho}.\nn
\eeqa
Some of the terms here are equal to zero, like:
$g f^{abc}G^{a}_{\mu\nu,\lambda}G^{b}_{\mu\rho,\lambda}\xi^{c}_{\nu\rho}$,
$g f^{abc}G^{a}_{\mu\lambda,\lambda}G^{b}_{\mu\nu,\nu}\xi^{c}_{\rho\rho}$
and $g f^{abc}G^{a}_{\mu\nu}G^{b}_{\mu\lambda}\xi^{c}_{\nu\lambda\rho\rho}$.
Amazingly all nonzero terms cancel each other.

In general one should construct all invariant quadratic forms for all tensor gauge fields.
A regular way of construction of {\it all invariant Lagrangians}
at every level is not known to the author, with exception of
the case of infinite series of particular invariants
presented in \cite{Savvidy:2005zm} by the formulas
(\ref{fulllagrangian1}) and (\ref{fulllagrangian2}).

\section{\it Extended Polygauge Groups }
We have three possible extensions of the Yang-Mills gauge group. The first
extension was defined by the equations (\ref{polygauge}) and
(\ref{generalgaugetransform}).  It allows definition of field strength
tensors (\ref{fieldstrengthparticular}) and (\ref{fieldstrengthgeneral})
transforming homogeneously with respect to the extended gauge
transformation (\ref{fieldstrenghparticular}) and (\ref{variationfieldstrengthgeneral}).

The second extension was also defined in \cite{Savvidy:2005zm} for the
higher-rank tensor fields $A_{\lambda_1 ...\lambda_s}$ which are totally symmetric tensors.
In that case the gauge transformation (\ref{polygauge}),
(\ref{generalgaugetransform}) should be modified in order
to respect the symmetry properties of the tensor fields.
Therefore the extended gauge transformation (\ref{polygauge}),
(\ref{generalgaugetransform})  was symmetrized over all indices as follows:
\beqa\label{polygaugesymmetric}
\tilde{\delta} A^{a}_{\mu} &=& ( \delta^{ab}\partial_{\mu}
+g f^{acb}A^{c}_{\mu})\xi^b ,~~~~~ \nonumber\\
(II)~~~~~~~\tilde{\delta} A^{a}_{\mu\nu} &=& \underline{( \delta^{ab}\partial_{\mu}
+  g f^{acb}A^{c}_{\mu})\xi^{b}_{\nu} }+ g f^{acb}A^{c}_{\mu\nu}\xi^{b},\nonumber\\
\tilde{\delta} A^{a}_{\mu\nu \lambda}& =& \underline{( \delta^{ab}\partial_{\mu}
+g f^{acb} A^{c}_{\mu})\xi^{b}_{\nu\lambda}} +
g f^{acb}(\underline{A^{c}_{\mu  \nu}\xi^{b}_{\lambda }} +A^{c}_{\mu\nu\lambda}\xi^{b}),\\
........&.&......................................,\nn
\eeqa
where
\beqa
\underline{( \delta^{ab}\partial_{\mu}
+  g f^{acb}A^{c}_{\mu})\xi^{b}_{\nu} } &=& ( \delta^{ab}\partial_{\mu}
+  g f^{acb}A^{c}_{\mu})\xi^{b}_{\nu} + ( \delta^{ab}\partial_{\nu}
+  g f^{acb}A^{c}_{\mu})\xi^{b}_{\mu}\nn\\
\underline{A^{c}_{\mu  \nu}\xi^{b}_{\lambda }} &=&
A^{c}_{\mu  \nu}\xi^{b}_{\lambda } +A^{c}_{\nu \lambda}\xi^{b}_{\mu  }
+A^{c}_{ \lambda  \mu}\xi^{b}_{\nu}\nn ,
\eeqa
and so on. One should also require that the
gauge parameters $\xi^{a}_{\lambda_1 ...\lambda_s}$ are totally symmetric
tensors. For example, in a matrix notation, we have
\beqa
\tilde{\delta}_{\xi}  A_{\mu\nu} = \partial_{\mu}\xi_{\nu} -i g[A_{\mu},\xi_{\nu}]
+\partial_{\nu}\xi_{\mu}  -i g[A_{\nu},\xi_{\mu}] -i g [A_{\mu\nu},\xi] .\nonumber
\eeqa
The commutator of the above gauge transformations acting on a rank-s tensor is
again a gauge transformation
\be
[\tilde{\delta}_{\eta},\tilde{\delta}_{\xi}] A_{\lambda_1 ...\lambda_s} = -i g~
\tilde{\delta}_{\zeta} A_{\lambda_1 ...\lambda_s},
\ee
and the gauge parameters form the same algebra as before (\ref{gaugealgebra}).
Therefore we concluded that the symmetrized gauge transformations
also form a closed algebraic structure \cite{Savvidy:2005zm}.

The third extension of the gauge transformations with double number of
gauge parameters can be defined as follows:
\beqa\label{doublepolygaugesymmetric}
\delta_{\xi} A^{a}_{\mu} &=& ( \delta^{ab}\partial_{\mu}
+g f^{acb}A^{c}_{\mu})\xi^b ,~~~~~ \nonumber\\
(III)~~~~~~~\delta_{\xi} A^{a}_{\mu\nu} &=&  ( \delta^{ab}\partial_{\mu}
+  g f^{acb}A^{c}_{\mu})\xi^{b}_{\nu} + g f^{acb}A^{c}_{\mu\nu}\xi^{b},\nonumber\\
\delta_{\eta} A^{a}_{\mu\nu} &=&  ( \delta^{ab}\partial_{\nu}
+  g f^{acb}A^{c}_{\nu})\eta^{b}_{\mu} + g f^{acb}A^{c}_{\mu\nu}\eta^{b},\\
........&.&......................................,\nn
\eeqa
and also forms a closed algebraic structure.

\section{{\it Conclusion}}
The transformations considered in the previous sections enlarge
the original algebra of Abelian local gauge transformations
found in \cite{Savvidy:2003fx} (expression (64) in \cite{Savvidy:2003fx}) to
a non-Abelian case
and unify into one multiplet particles with arbitrary spins
and with linearly growing multiplicity.
As we have seen, this leads to a natural generalization of the Yang-Mills theory.
The extended non-Abelian gauge transformations
defined for the tensor gauge fields led to the construction of the
appropriate field strength tensors and of the invariant Lagrangians.
The proposed extension may lead  to a natural inclusion of the standard
theory of fundamental forces into a larger theory in which standard
particles (vector gauge bosons, leptons and quarks) represent a
low-spin subgroup of an enlarged family of particles with higher spins.

As an example of an extended gauge field theory with infinite many gauge fields,
this theory {\it can be associated} with
the field theory of the tensionless strings, because in
our generalization of the non-Abelian Yang-Mills theory we essentially
used the symmetry group which appears as symmetry of the ground state
wave function of the tensionless strings
\cite{Savvidy:2003fx,Savvidy:dv,Savvidy:2005fe}.
Nevertheless I do not know how to
derive it directly from tensionless strings, therefore one can not claim that they are
indeed identical. The main reason is that the above construction, which is purely
field-theoretical, has a great advantage of being well defined on and off the mass-shell,
while the string-theoretical constructions have not been yet developed to
the same level, because the corresponding vertex operators are well defined
only on the mass-shell \cite{Savvidy:2005fe}.
The tensor gauge field theory could probably be
a genuine tensionless string field theory because of the common  symmetry
group, and it would be useful to
understand, whether the string theory can fully reproduce this result.
Discussion of the tensionless strings and related questions can also be
found in \cite{Edgren:2005gq,Turok:2004gb,
Engquist:2005yt,Mourad:2005rt,
Bekaert:2005vh,Brink:2005wh,Gamboa:2004cv,Bonelli:2004ve,Bakas:2004jq,
Bredthauer:2004kv,Gamboa:2003fy,Chagas-Filho:2003wv,Bianchi:2003wx,
Gabrielli:1990ay,Gabrielli:1999xt,Baez:2002jn,Castro:2004hi,Bengtsson:2004cd}.

This work was partially supported by the EEC Grant no. HPRN-CT-1999-00161 and
EEC Grant no. MRTN-CT-2004-005616.

\vfill

\begin{thebibliography}{99}

\bibitem{yang} C.N.Yang and R.L.Mills. "Conservation of Isotopic
Spin and Isotopic Gauge Invariance". Phys.\ Rev.\ {\bf 96} (1954) 191

\bibitem{yukawa}H.Yukawa.  "On the Interaction
of Elementary Particles" I, Proc.\ Phys.\ Math.\ Soc.\ Japan {\bf 17} (1935) 48

\bibitem{Schwinger:1957em}
J.~S.~Schwinger,
``A Theory Of The Fundamental Interactions,''
Annals Phys.\  {\bf 2} (1957) 407.


\bibitem{Glashow:1961tr}
S.~L.~Glashow,
``Partial Symmetries Of Weak Interactions,''
Nucl.\ Phys.\  {\bf 22} (1961) 579.

\bibitem{salam}A.Salam, In Elementary Particle Theory: Relativistic Groups
and Analyticity (Nobel Symposium No.8), edited by N.Svartholm,
(Almqvist and Wiksell, Stockholm,1968)

\bibitem{weinberg}S.Weinberg. A model of Leptons.\  Phys.\ Rev.\  Lett.\ {\bf 19} (1967) 1264

\bibitem{Feynman:1963ax}
  R.~P.~Feynman,
  ``Quantum Theory Of Gravitation,''
  Acta Phys.\ Polon.\  {\bf 24} (1963) 697.

\bibitem{Faddeev:1967fc}
  L.~D.~Faddeev and V.~N.~Popov,
  ``Feynman Diagrams For The Yang-Mills Field,''
  Phys.\ Lett.\ B {\bf 25} (1967) 29;~
    ``Perturbation Theory For Gauge Invariant Fields,''
FERMILAB-PUB-72-057-T

\bibitem{DeWitt:1967ub}
  B.~S.~DeWitt,
  ``Quantum Theory Of Gravity. II,III,''
  Phys.\ Rev.\  {\bf 162} (1967) 1195,1239.


\bibitem{Mandelstam:1968hz}
  S.~Mandelstam,
  ``Feynman Rules For Electromagnetic And Yang-Mills Fields From The Gauge
  Independent Field Theoretic Formalism,''
  Phys.\ Rev.\  {\bf 175} (1968) 1580.

\bibitem{Slavnov:1970tk}
  A.~A.~Slavnov and L.~D.~Faddeev,
  ``Massless And Massive Yang-Mills Field. (In Russian),''
  Theor.\ Math.\ Phys.\  {\bf 3} (1970) 312
  [Teor.\ Mat.\ Fiz.\  {\bf 3} (1970) 18].


\bibitem{'tHooft:1971fh}
  G.~'t Hooft,
  ``Renormalization Of Massless Yang-Mills Fields,''
  Nucl.\ Phys.\ B {\bf 33} (1971) 173.

\bibitem{'tHooft:1971rn}
  G.~'t Hooft,
  ``Renormalizable Lagrangians For Massive Yang-Mills Fields,''
  Nucl.\ Phys.\ B {\bf 35} (1971) 167.

\bibitem{'tHooft:1972fi}
  G.~'t Hooft and M.~J.~G.~Veltman,
  ``Regularization And Renormalization Of Gauge Fields,''
  Nucl.\ Phys.\ B {\bf 44} (1972) 189.

\bibitem{Lee:1971kj}
  B.~W.~Lee,
  ``Renormalizable Massive Vector Meson Theory. Perturbation Theory Of The
  Higgs Phenomenon,''
  Phys.\ Rev.\ D {\bf 5} (1972) 823.

\bibitem{Lee:1972fj}
  B.~W.~Lee and J.~Zinn-Justin,
  ``Spontaneously Broken Gauge Symmetries. I,II,III.''
  Phys.\ Rev.\ D {\bf 5} (1972) 3121,3137,3155.

\bibitem{Slavnov:1972fg}
  A.~A.~Slavnov,
  ``Ward Identities In Gauge Theories,''
  Theor.\ Math.\ Phys.\  {\bf 10} (1972) 99
  [Teor.\ Mat.\ Fiz.\  {\bf 10} (1972) 153].

\bibitem{Taylor:1971ff}
  J.~C.~Taylor,
  ``Ward Identities And Charge Renormalization Of The Yang-Mills Field,''
  Nucl.\ Phys.\ B {\bf 33} (1971) 436.

\bibitem{Gross:1973id}
  D.~J.~Gross and F.~Wilczek,
   ``Ultraviolet Behavior Of Non-Abelian Gauge Theories,''
  %
  Phys.\ Rev.\ Lett.\  {\bf 30}, 1343 (1973).

\bibitem{Politzer:1973fx}
  H.~D.~Politzer,
   ``Reliable Perturbative Results For Strong Interactions?,''
  Phys.\ Rev.\ Lett.\  {\bf 30}, 1346 (1973).

\bibitem{Fritzsch:1973pi}
  H.~Fritzsch, M.~Gell-Mann and H.~Leutwyler,
  ``Advantages Of The Color Octet Gluon Picture,''
  Phys.\ Lett.\ B {\bf 47} (1973) 365.

\bibitem{Savvidy:1977as}
  G.~K.~Savvidy,
  ``Infrared Instability Of The Vacuum State Of Gauge Theories And Asymptotic
  Freedom,''
  Phys.\ Lett.\ B {\bf 71} (1977) 133.

\bibitem{Savvidy:2005zm}
G.~Savvidy,
``Generalization of Yang-Mills theory: Non-Abelian tensor gauge fields and
higher-spin extension of standard model,''
arXiv:hep-th/0505033.

\bibitem{fock} V.Fock. \"Uber die invariante Form der Wellen- und der
Bewegungsgleichungen f\"ur einen geladenen Massenpunkt. Z. f\"ur Physik 39 (1927) 226

\bibitem{klein} O.Klein. Quantentheorie und f\"unfdimensionale Relativit\"atstheorie.
 Z. f\"ur Physik 37 (1926) 895

\bibitem{london} F.London. Quantenmechanische Deutung der Theorie von Weyl.
Z. f\"ur Physik 42 (1927) 375

\bibitem{weyl} H.Weyl. "Elektron und Gravitation" Z. f\"ur Physik 56 (1929) 330;\\
H.Weyl. Gruppentheorie und Quantenmechanik. Ch.IV.  Leipzig,1928

\bibitem{pauli}W.Pauli und W.Heisenberg. " Zur Quantentheorie
der Wellenfelder. II" Z. f\"ur Physik 59 (1930) 168;\\
W.Pauli. Relativistic Field Theories of Elementary
Particles. \ Rev.\ Mod.\ Phys.\ {\bf 13} (1941) 203

\bibitem{chern}S.S.Chern. {\it Topics in Defferential Geometry},
Ch. III "Theory of Connections"
(The Institute for Advanced Study, Princeton, 1951)


\bibitem{majorana} E.Majorana. Teoria Relativistica di Particelle con Momento
Intrinseco Arbitrario. Nuovo Cimento {\bf 9} (1932) 335

\bibitem{dirac} P.A.M.Dirac. Relativistic wave equations.
Proc.\ Roy.\ Soc.\ {\bf A155} (1936) 447;\\
Unitary Representation of the Lorentz Group.
Proc.\ Roy.\ Soc.\ {\bf A183} (1944) 284.


\bibitem{fierzpauli}M.~Fierz. Helv.\ Phys.\ Acta.\  {\bf 12} (1939) 3.\\
M.~Fierz and W.~Pauli.  On Relativistic Wave Equations for
Particles of Arbitrary Spin in an Electromagnetic Field. Proc.\ Roy.\ Soc.\  {\bf A173} (1939) 211.

\bibitem{wigner}E.~Wigner. On Unitary Representations of the Inhomogeneous
Lorentz Group.  Ann. Math. {\bf 40} (1939) 149.


\bibitem{rarita}W.~Rarita and J.~Schwinger. On a Theory of Particles with
Half-Integral Spin.  Phys.\ Rev. {\bf 60} (1941) 61

\bibitem{yukawa1}H.Yukawa.  "Quantum Theory of Non-Local Fields.
Part I. Free Fields" Phys.\ Rev. {\bf 77} (1950) 219\\
M.~Fierz. "Non-Local Fields" Phys.\ Rev. {\bf 78} (1950) 184

\bibitem{schwinger}J.Schwinger. {\it Particles, Sourses, and Fields}
(Addison-Wesley, Reading, MA, 1970)

\bibitem{Weinberg:1964cn}
S.~Weinberg,
``Feynman Rules For Any Spin,''
Phys.\ Rev.\  {\bf 133} (1964) B1318.

\bibitem{Weinberg:1964ev}
S.~Weinberg,
``Feynman Rules For Any Spin. 2: Massless Particles,''
Phys.\ Rev.\  {\bf 134} (1964) B882.

\bibitem{Weinberg:1964ew}
S.~Weinberg,
``Photons And Gravitons In S Matrix Theory: Derivation Of Charge Conservation
And Equality Of Gravitational And Inertial Mass,''
Phys.\ Rev.\  {\bf 135} (1964) B1049.


\bibitem{chang}S.~J.~Chang. Lagrange Formulation for Systems with Higher Spin.
Phys.Rev. {\bf 161} (1967) 1308

\bibitem{singh}L.~P.~S.~Singh and C.~R.~Hagen. Lagrangian formulation for
arbitrary spin. I. The boson case.
Phys.\ Rev.\ {\bf D9} (1974) 898

\bibitem{singh1}L.~P.~S.~Singh and C.~R.~Hagen. Lagrangian formulation for
arbitrary spin. II. The fermion case.
Phys.\ Rev.\ {\bf D9} (1974) 898, 910

\bibitem{fronsdal}C.Fronsdal.
Massless fields with integer spin,
Phys.Rev. {\bf D18} (1978) 3624

\bibitem{fronsdal1}J.Fang and C.Fronsdal.
Massless fields with half-integral spin,
Phys.\ Rev.\ {\bf D18} (1978) 3630


\bibitem{Gupta} N.S.Gupta. "Gravitation and Electromagnetism".
Phys.\ Rev.\  {\bf 96}, (1954) 1683 .

\bibitem{kraichnan} R.H.Kraichnan. "Special-relativistic derivation
of generally covariant gravitation theory".
Phys.\ Rev.\  {\bf 98}, (1955) 1118.

\bibitem{thirring} W.E.Thirring. "An alternative approach to the
theory of gravitation". Ann.\ Phys.\  {\bf 16}, (1961) 96.

\bibitem{feynman} R.P.Feynman. "Feynman Lecture on Gravitation".
Westview Press 2002.

\bibitem{deser} S.Deser. "Self-interaction and gauge invariance".
Gen.\ Rel.\ Grav.\ {\bf 1}, (1970) 9.

\bibitem{fronsdal2}J.Fang and C.Fronsdal.
"Deformation of gauge groups. Gravitation".
J.\ Math.\ Phys.\ {\bf 20} (1979) 2264

\bibitem{witten}
S.Weinberg and E.Witten,
``Limits on massless particles,''
Phys.\ Lett.\ B {\bf 96} (1980) 59.


\bibitem{deser1}
C.~Aragone and S.~Deser,
``Constraints on gravitationally coupled tensor fields,''
Nuovo.\ Cimento.\ {\bf 3}A (1971) 709;\\
C.Aragone and S.Deser,
``Consistancy problems of hypergavity,''
Phys.\ Lett.\ B {\bf 86} (1979) 161.

\bibitem{berends}
F.~A.~Berends, G.~J.~H~Burgers and H.~Van Dam,
``On the Theoretical problems in Constructing Interactions Involving
Higher-Spin Massless Particles,''
Nucl.\ Phys.\ B {\bf 260} (1985) 295.

\bibitem{dewit}
B.deWit, F.A.~Berends, J.W.van Halten and P.Nieuwenhuizen,
``On Spin-5/2 gauge fields,''
J.\ Phys.\ A: Math.\ Gen.\  {\bf 16} (1983) 543.


\bibitem{vasiliev}M.A.Vasiliev, "Progress in Higher Spin Gauge Theories"
hep-th/0104246;
M.A.Vasiliev et. al. "Nonlinear Higher Spin Theories in Various Domensions"
hep-th/0503128

\bibitem{Sezgin:2001zs}
E.~Sezgin and P.~Sundell,
``Holography in 4D (Super) Higher Spin Theories and
a Test via Cubic Scalar Couplings,'' hep-th/0305040

\bibitem{Sagnotti:2005ns}
  A.~Sagnotti, E.~Sezgin and P.~Sundell,
  ``On higher spins with a strong Sp(2,R) condition,''
  arXiv:hep-th/0501156.

\bibitem{Bengtsson:1983pd}
A.~K.~Bengtsson, I.~Bengtsson and L.~Brink,
``Cubic Interaction Terms For Arbitrary Spin,''
Nucl.\ Phys.\ B {\bf 227} (1983) 31.

\bibitem{Bengtsson:1983pg}
A.~K.~Bengtsson, I.~Bengtsson and L.~Brink,
``Cubic Interaction Terms For Arbitrarily Extended Supermultiplets,''
Nucl.\ Phys.\ B {\bf 227} (1983) 41.



\bibitem{Savvidy:dv}
G.~K.~Savvidy, Conformal Invariant Tensionless Strings,
Phys. Lett.\ B {\bf 552} (2003) 72.


\bibitem{Savvidy:2003fx}
G.~K.~Savvidy,~
``Tensionless strings: Physical Fock space and higher spin fields,''
Int.\ J.\ Mod.\ Phys.\ A {\bf 19},  (2004) 3171-3194.


\bibitem{Savvidy:2005fe}
G.~Savvidy,
``Tensionless strings, correspondence with SO(D,D) sigma model,''
Phys.\ Lett.\ B {\bf 615} (2005) 285.


\bibitem{Edgren:2005gq}
  L.~Edgren, R.~Marnelius and P.~Salomonson,
  ``Infinite spin particles,''
  JHEP {\bf 0505} (2005) 002
  [arXiv:hep-th/0503136].

\bibitem{Turok:2004gb}
  N.~Turok, M.~Perry and P.~J.~Steinhardt,
  ``M theory model of a big crunch / big bang transition,''
  Phys.\ Rev.\ D {\bf 70} (2004) 106004
  [arXiv:hep-th/0408083].

\bibitem{Engquist:2005yt}
  J.~Engquist and P.~Sundell,
  ``Brane partons and singleton strings,''
  arXiv:hep-th/0508124.

\bibitem{Mourad:2005rt}
  J.~Mourad,
  ``Continuous spin particles from a string theory,''
  arXiv:hep-th/0504118.

\bibitem{Bekaert:2005vh}
  X.~Bekaert, S.~Cnockaert, C.~Iazeolla and M.~A.~Vasiliev,
  ``Nonlinear higher spin theories in various dimensions,''
  arXiv:hep-th/0503128.

\bibitem{Brink:2005wh}
  L.~Brink,
  ``Particle physics as representations of the Poincare algebra,''
  arXiv:hep-th/0503035.

\bibitem{Gamboa:2004cv}
  J.~Gamboa, M.~Loewe and F.~Mendez,
  ``Quantum theory of tensionless noncommutative p-branes,''
  Phys.\ Rev.\ D {\bf 70} (2004) 106006.

\bibitem{Bonelli:2004ve}
  G.~Bonelli,
  ``On the boundary gauge dual of closed tensionless free strings in AdS,''
  JHEP {\bf 0411} (2004) 059
  [arXiv:hep-th/0407144].

\bibitem{Bakas:2004jq}
  I.~Bakas and C.~Sourdis,
  ``On the tensionless limit of gauged WZW models,''
  JHEP {\bf 0406} (2004) 049
  [arXiv:hep-th/0403165].

\bibitem{Bredthauer:2004kv}
  A.~Bredthauer, U.~Lindstrom, J.~Persson and L.~Wulff,
  `Type IIB tensionless superstrings in a pp-wave background,''
  JHEP {\bf 0402} (2004) 051
  [arXiv:hep-th/0401159].

\bibitem{Gamboa:2003fy}
  J.~Gamboa and F.~Mendez,
  ``Statistical quantum mechanics of many universes,''
  arXiv:hep-th/0304116.

\bibitem{Chagas-Filho:2003wv}
  W.~F.~Chagas-Filho,
  ``Symmetries in particle and string theories,''
  arXiv:hep-th/0309219.

\bibitem{Bianchi:2003wx}
  M.~Bianchi, J.~F.~Morales and H.~Samtleben,
  ``On stringy AdS(5)x $ S^5$ and higher spin holography,''
  JHEP {\bf 0307} (2003) 062
  [arXiv:hep-th/0305052].


\bibitem{Gabrielli:1990ay}
  E.~Gabrielli,
  ``Extended Pure Yang-Mills Gauge Theories With Scalar And Tensor Gauge
  Phys.\ Lett.\ B {\bf 258} (1991) 151.


\bibitem{Gabrielli:1999xt}
  E.~Gabrielli,
  ``Extended gauge theories in Euclidean space with higher spin fields,''
  Annals Phys.\  {\bf 287} (2001) 229
  [arXiv:hep-th/9909117].

\bibitem{Baez:2002jn}
  J.~C.~Baez,
  ``Higher Yang-Mills theory,''
  arXiv:hep-th/0206130.

\bibitem{Castro:2004hi}
  C.~Castro,
  ``Generalized p-forms electrodynamics in Clifford spaces,''
  Mod.\ Phys.\ Lett.\ A {\bf 19} (2004) 19.

\bibitem{Bengtsson:2004cd}
  A.~K.~H.~Bengtsson,
  J.\ Math.\ Phys.\  {\bf 46} (2005) 042312
  [arXiv:hep-th/0403267].





\end{thebibliography}
\end{document}